\documentclass[aps,prb,twocolumn,amsmath,amssymb,longbibliography,groupedaddress]{revtex4-2} 

\usepackage{epsfig}
\usepackage{subfigure}
\usepackage{graphicx}
\usepackage{amsfonts}
\usepackage[figuresright]{rotating}
\usepackage{amssymb}
\usepackage{amsmath}
\usepackage{psfrag}
\usepackage{esint} 
\usepackage{bm}
\usepackage[colorlinks,linkcolor=blue,anchorcolor=blue,citecolor=blue,urlcolor=blue]{hyperref}
\usepackage[version=4]{mhchem}
\usepackage{multirow}
\usepackage{siunitx}
\usepackage{xcolor}

\def\be{\begin{equation}} \def\ee{\end{equation}}
\def\bea{\begin{eqnarray}} \def\eea{\end{eqnarray}}

\def\k{{\bf k}}

\def\p{{\bf p}}

\def\bpm{\begin{pmatrix}} \def\epm{\end{pmatrix}}

\newcommand\pmat[1]{\begin{pmatrix}#1\end{pmatrix}}

\makeatletter
\newcommand*{\balancecolsandclearpage}{%
  \close@column@grid
  \clearpage
}
\makeatother

\begin{document}

\title{Monopole Superconductivity in Magnetically Doped Cd$_3$As$_2$}

\author{Eric Bobrow}
\address{Institute for Quantum Matter and Department of Physics and Astronomy, Johns Hopkins University, Baltimore, Maryland 21218, USA}
\author{Yi Li}
\address{Institute for Quantum Matter and Department of Physics and Astronomy, Johns Hopkins University, Baltimore, Maryland 21218, USA}

\begin{abstract}
    When superconducting pairing occurs between Fermi surfaces with different Chern numbers, the Cooper pairs possess nontrivial pair Berry phase, which enforces pairing gap nodes. The resulting pairing order is further distinguished from the familiar $s$-, $p$-, and $d$-wave pairing orders by a nonzero pair monopole charge and is described by monopole harmonics. 
    To date, this exotic monopole pairing order is yet to be achieved experimentally. 
    We therefore study the magnetically doped Dirac semimetal Cd$_3$As$_2$ as a candidate material for realizing monopole superconductivity with pair monopole charges $q_p=1$ or $2$, depending on the chemical potential. 
    For each case of pair monopole charge, we explore representatives of uniform pairing orders in all allowed irreducible representations of the $C_{4h}$ symmetry of magnetically doped Cd$_3$As$_2$. 
    We demonstrate the distinctions in the monopole analogs of different higher partial wave superconducting orders that result from the combination of topological Fermi surfaces of higher Chern number and crystalline symmetry. 
    In all cases, the patterns of the superconducting phase winding around a Fermi surface are constrained by the topologically invariant total winding number, which depends only on the pair monopole charge and requires nodes in the $q_p\neq 0$ superconducting order. 
    Our work can guide further experimental investigation of monopole orders in the topological semimetal Cd$_3$As$_2$ and can be generalized to other materials with Fermi surfaces of higher Chern number, which enforce monopole pairing order in higher partial waves, such as the monopole analog of $f$-wave pairing.
\end{abstract}

\maketitle

Bloch states residing on topological Fermi surfaces have been found to give rise to remarkable physical properties, from intrinsic anomalous Hall effects to topological states localized at boundaries
\cite{Karplus1954,Haldane1988,Jungwirth2002,Onoda2002,Haldane2004,Chang2013,Nagaosa2010,Wan2011,Yang2011,Xu2015,Lv2015,Moll2016,Nakatsuji2015}. 
In the presence of broken time-reversal or inversion symmetry, the single-particle Bloch state on a topological Fermi surface possesses nonzero Berry curvature when the phase of the state is obstructed from being smoothly defined over the Fermi surface, as characterized by a nonzero Chern number.

In proximity to an $s$-wave superconductor, the Fermi surface topology further leads to a zoo of topological superconductor candidates  \cite{Fu2008,Lutchyn2010,Li2018,Sun2019,Li2020,Sun2020,Yu2021} in the weak-coupling regime where the phases of the Bogoliubov-de Gennes (BdG) single-quasiparticle states fail to be globally well defined around the Fermi surface. 
When superconducting pairing occurs between opposite momentum states from a single topological Fermi pocket, the pairing order can exhibit an effective unconventional chiral $p$-wave pairing symmetry and Majorana states localized at the surface or edges of the system \cite{Read2000,Fu2008,Lutchyn2010,Volovik2009,Chung2009,Qi2010,Alicea2010,Alicea2012}. 
In contrast, when pairing occurs between Fermi pockets of different Chern numbers, not only the BdG quasiparticle states but also the pairing order are topologically obstructed, and the phase of the pairing order cannot be globally well defined over the Fermi surface.
The obstructed pairing order can be understood in terms of the two-particle Berry phase of a Cooper pair \cite{Murakami2003a}, which necessitates a description of the topology and symmetry of the superconducting pairing order in terms of monopole harmonics rather than spherical harmonics, or monopole superconductivity \cite{Li2018,Sun2019, Munoz2020,Li2020}. 
The notion of monopole pairing order has been generalized to monopole density-wave order in the particle-hole channel \cite{Bobrow2020} and to non-Abelian topological obstructions of the superconducting order characterized by $\mathbb{Z}_2$ \cite{Sun2020} and Euler \cite{Yu2021} indices in the presence of time reversal and combined time reversal and inversion symmetry. 

Recent work has revealed that the Dirac semimetal Cd$_3$As$_2$ is a potential platform to realize monopole superconductivity when time reversal symmetry is broken. 
The Dirac semimetal Cd$_3$As$_2$ has been experimentally found to exhibit pressure- or proximity-induced  superconductivity \cite{Wang2016d, Aggarwal2016, Huang2019, Cuozzo2022}, 
with the magnetically doped material being actively pursued \cite{Xiao2022}. 
Recent first principle calculations of magnetically doped Cd$_3$As$_2$ have shown a rich Weyl node structure that supports topological Fermi surfaces with Chern numbers $\pm 1$ and $\pm 2$  \cite{Baidya2020}. 
The Zeeman exchange field from magnetic dopants breaks time reversal symmetry while preserving the parity of the low-temperature phase of Cd$_3$As$_2$ \cite{Ali2014b}. 
Therefore, inter-Fermi surface pairing occurs between parity-related Fermi pockets with opposite Chern numbers, opening up the opportunity for realizing the monopole superconducting order. 
The monopole charge of the Cooper pair formed between Fermi pockets with Chern numbers $\pm C$ is topological invariant and fundamentally influences the pairing, placing a lower bound on the pairing partial-wave channel and requiring a description in terms of monopole harmonics.

\begin{figure*}
\subfigure[]{\centering \epsfig{file=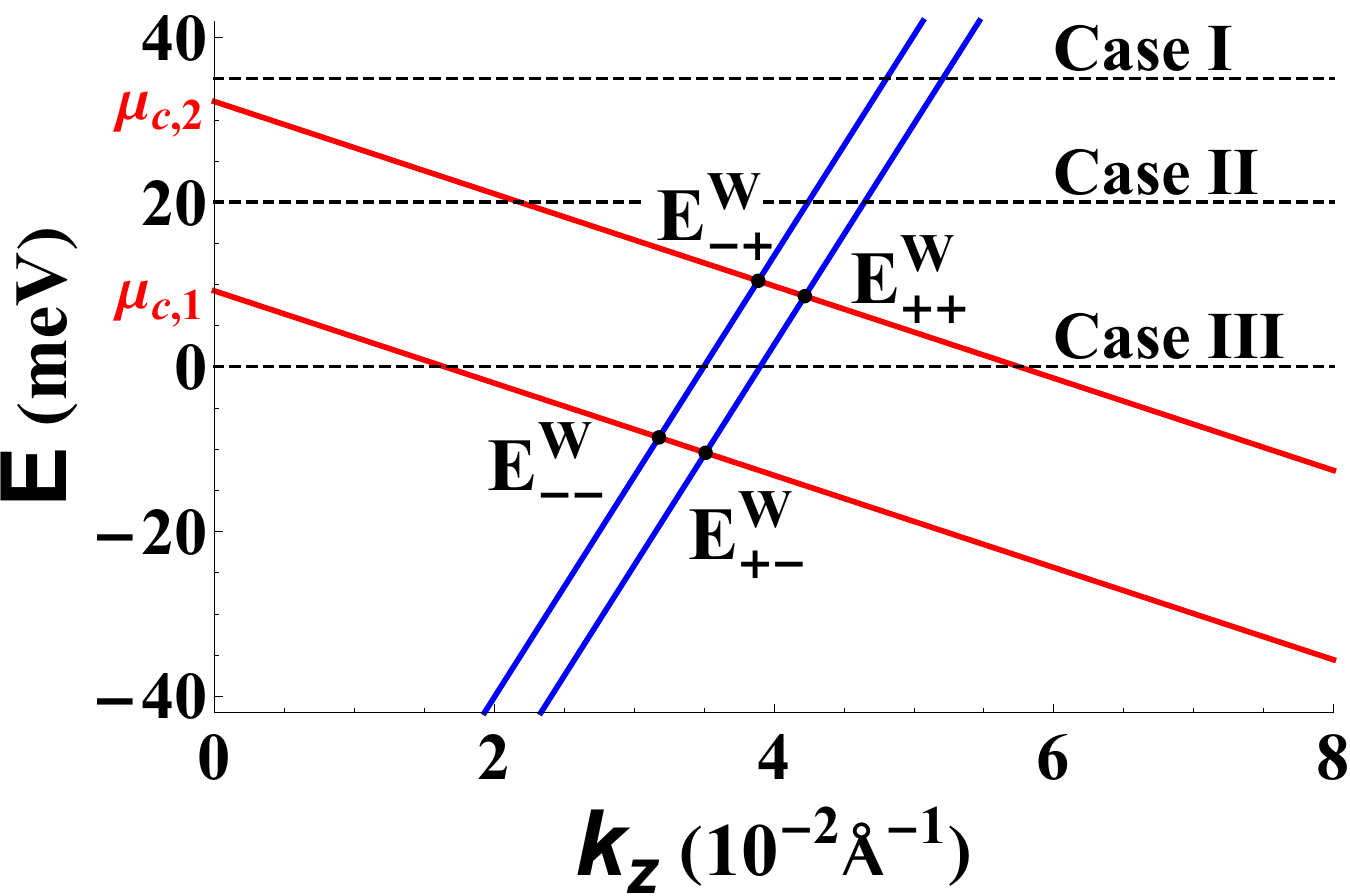, width=0.3\linewidth}
\label{fig:kzspect}}
\subfigure[]{\centering \epsfig{file=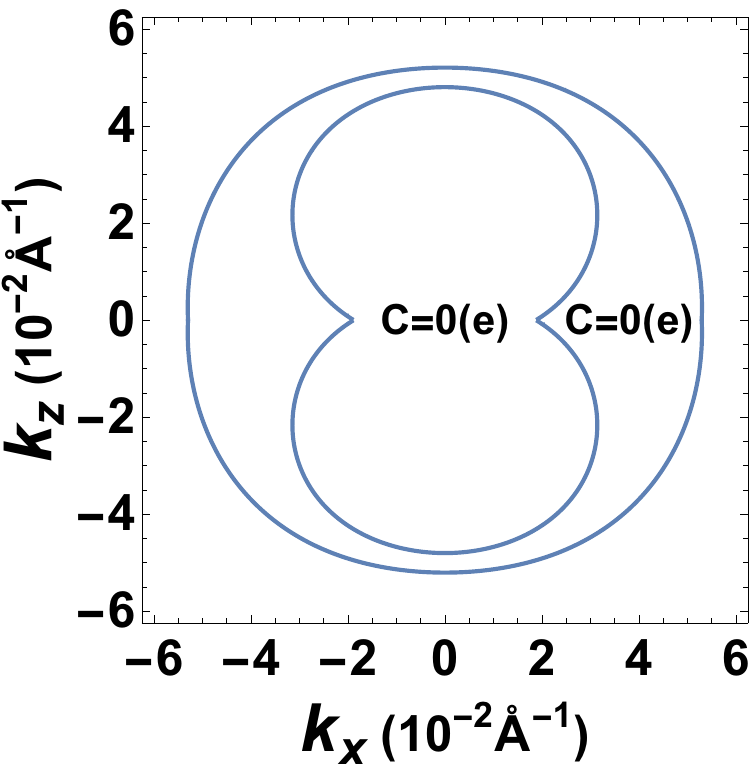, width=0.207\linewidth}
\label{fig:mu0035BandFS}}
\subfigure[]{\centering \epsfig{file=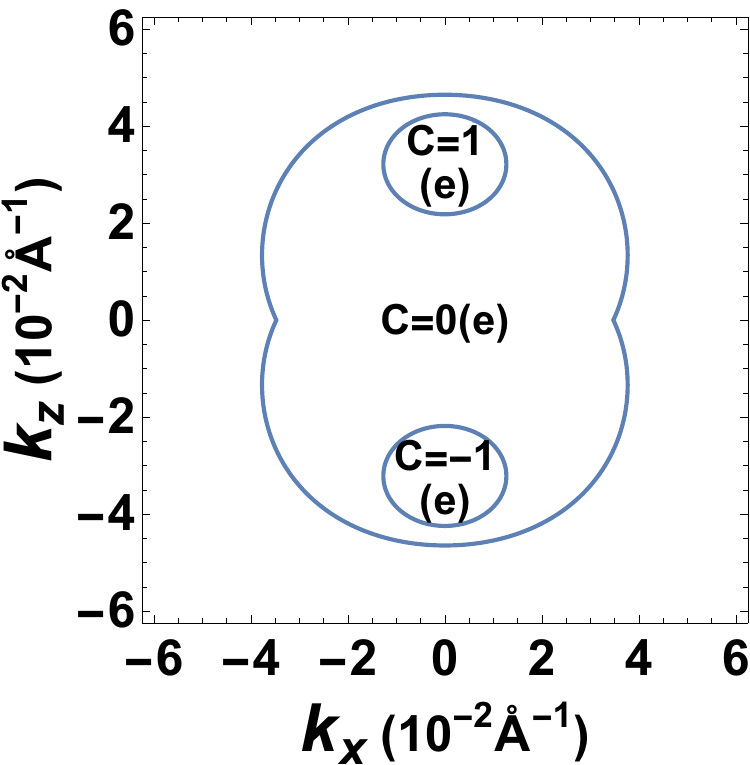, width=0.207\linewidth}
\label{fig:mu002BandFS}}
\subfigure[]{\centering \epsfig{file=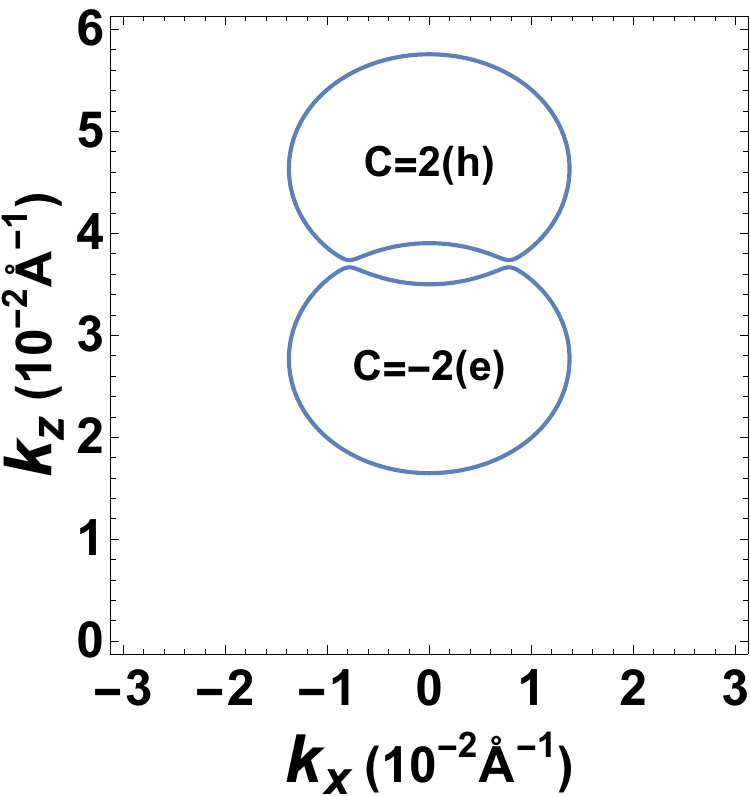, width=0.2\linewidth}
\label{fig:mu0BandFS}}
\caption{($a$) Bands of ${\cal{H}}(\k)$ for $k_z \ge 0$ with $s$- and $p$-orbital bands in blue and red, respectively. 
The three dashed lines indicate Cases I, II and III at $\mu=35$, $20$, and $0\si{meV}$ respectively with the corresponding Fermi pockets at $k_y=0$ shown in ($b$), ($c$), and ($d$). 
Here, ``C" denotes the Chern number and ``e" (``h") indicates an electron (hole) Fermi pocket. 
In ($d$), a pair of Fermi pockets at $k_z < 0$ related by parity to those shown is omitted.
}
\label{fig:BandFS}
\end{figure*}

In this paper, we explore different uniform superconducting pairing orders allowed by symmetry in magnetically doped Cd$_3$As$_2$. 
In three topologically distinct cases, pairing between Fermi pockets with opposite Chern numbers $C=0$, $\pm 1$, and $\pm 2$
at chemical potentials of $35$, $20$, and $0$meV above the zero-field Dirac node energy, we demonstrate different classes of pairing order distinguishable by the topological invariant $\nu_{tot}=2C$. This invariant is the total momentum-space vorticity that can be identified with the 
winding of the superconducting phase over a paired Fermi pocket \cite{Murakami2003a,Li2018}, and monopole pairing order results for the latter two cases, where the topological invariant is nontrivial. 

The crystalline symmetry in magnetically doped Cd$_3$As$_2$ further enriches the superconducting order and influences the nodal structure of monopole pairing. Within the same topological class, the pairing order can exhibit gap node and phase winding patterns with different symmetries. For the monopole pairing between the high Chern number $C=\pm 2$ Fermi pockets, we find nonvanishing monopole pairing order in the $A_u$, $B_u$, $^1E_u$, and $^2E_u$ representations of the $C_{4h}$ point group symmetry, with different phase winding patterns in this case subject to the same topological invariant $\nu_{tot} = 4$.


\textit{Model of superconducting magnetically doped  Cd$_3$As$_2$.} 
Magnetically doped Cd$_3$As$_2$ provides a useful platform for material realization of monopole superconducting order. 
The inversion-symmetric low temperature phase of Cd$_3$As$_2$ features Dirac nodes at $\pm\k_D$ \cite{Ali2014b} that can be described by a $\k\cdot\p$ model \cite{Wang2013} (see also Sec. IA of the Supplemental Material (S. M.)). In the presence of magnetic dopants, represented by a Zeeman exchange field along the $z$ direction, the symmetry is reduced from point group $D_{4h}$ to $C_{4h}$, and each Dirac node splits into multiple inversion-related Weyl nodes \cite{Baidya2020}. 
Hence, when the system is doped near the Weyl nodes, the system features disconnected inversion-related Fermi pockets with opposite Chern numbers. As the topological Fermi pockets separately lack inversion symmetry about their centers, superconducting pairing between inversion-related Fermi pockets is energetically preferred to intra-Fermi pocket pairing, leading to the possibility of monopole superconductivity \cite{Li2018}. 

In mean-field theory, the zero center-of-mass momentum pairing 
can be described by the BdG Hamiltonian
\begin{equation}
    {\cal{H}}_{BdG}(\k) = \pmat{{\cal{H}}(\k) & \hat{\Delta}(\k) \\ \hat{\Delta}^\dagger(\k) & -\cal{H}^*(-\k)},
\end{equation}
where ${\cal{H}}(\k)$ is a four-band model of magnetically doped Cd$_3$As$_2$ and the superconducting pairing matrix $\hat{\Delta}(\k)$ is constrained by the $C_{4h}$ point group symmetry.


To capture the topologically distinct Fermi pockets with a simple model, $\mathcal{H}(\k)$ describes the Weyl nodes split from the zero-field Dirac points $\pm \k_D = (0, 0, \pm k_D)$ with linearized dispersion in the $k_z$ direction \cite{Baidya2020}. Near $+\k_D$, 
\begin{equation}
\begin{aligned}
   &\mathcal{H}_+ (\k) = (\epsilon_D - \mu) I +\\
    &\begin{pmatrix}
        v_s\tilde{k}_z + \beta_s h & A k_+ & 0 & G k_-^2 \\
        A k_- & -v_p\tilde{k}_z + \beta_p h & G k_-^2 & 0 \\
        0 & G k_+^2 & v_s\tilde{k}_z - \beta_s h & -A k_- \\
        G k_+^2 & 0 & -A k_+ & -v_p \tilde{k}_z - \beta_p h
    \end{pmatrix}
    \label{eq:H}
\end{aligned}
\end{equation}
in the spin-orbit coupled basis $\{|S_{J=\frac{1}{2}}, J_z=\frac{1}{2}\rangle, |P_{\frac{3}{2}}, \frac{3}{2}\rangle, |S_{\frac{1}{2}}, -\frac{1}{2}\rangle, |P_{\frac{3}{2}}, -\frac{3}{2}\rangle\}$ of Cd $5s$ and As $4p$ orbitals.
To maintain inversion symmetry, $\cal{H}(\k)$ can be defined piecewise as $\mathcal{H}(\k) = \mathcal{H}_{+}(\k)$ $(\mathcal{H}_{-}(\k))$ for $k_z >0$ ($<0$) related by $\mathcal{H}_-(\k) = D(\Pi){\mathcal{H}}_{+}(-\k) D(\Pi)^\dagger$ with $D(\Pi) = I \otimes \tau_z$ the representation of parity in terms of tensor products of Pauli matrices in the spin-orbit coupled basis.
In Eq. \eqref{eq:H}, $\tilde{\k} = \k - \k_D$ is the momentum relative to the zero-field Dirac node, $k_\pm = k_x \pm i k_y$,  
$\beta_s$ and $\beta_p$ are the Zeeman couplings for the $s$ and $p$ bands under the Zeeman field of magnitude $h$ along the $z$ direction, and $v_s$ and $v_p$ are the Fermi velocities along $k_z$ near $\k_D$. 
The zero-field Dirac node energy $\epsilon_D$ is set to zero, and the remaining parameter values, calculated from density functional theory in Ref. \cite{Baidya2020}, are included for convenience in S. M. Sec. IB. 
The band dispersion along the positive $k_z$ axis is shown in Fig. \ref{fig:kzspect} and features four Weyl nodes at $\k^W_{s_1, s_2} = \k_D + (0,0,\frac{s_1 \beta_s + s_2 \beta_p}{v_s + v_p}h)$ with energies $E^W_{s_1, s_2} = \frac{-s_1 v_p\beta_s + s_2 v_s \beta_p}{v_s + v_p}h$, labeled by signs $s_1, s_2 = \pm 1$. 

Distinct superconducting pairing orders can arise between Fermi pockets with different Chern numbers, and the pairing order thus depends crucially on the band topology. 
As this model exhibits Fermi pockets with Chern numbers $0$, $\pm1$, and $\pm2$ at different chemical potentials, we focus on the superconducting pairing in the representative cases $\mu = 35$, $20$, and $0$meV, referred to respectively as Cases I, II, and III  as labeled in Fig. \ref{fig:kzspect}. 
In each case, the topological structure of the Fermi surface is determined by the chemical potential relative to the Weyl node energies as well as the band energies at the $\Gamma$ point. The $\Gamma$ point band energies determine critical chemical potentials where inversion-related Fermi pockets merge into a single parity-symmetric Fermi pocket, $\mu_{c,1} = 9.2$meV and $\mu_{c, 2} = 32$meV, as well as  $-94$meV and $-105$meV at $\mu < 0$. 
For Case I, where $\mu > \mu_{c,2}$, the Fermi pockets are inversion symmetric and topologically trivial, as shown in Fig. \ref{fig:mu0035BandFS}, and intra-Fermi pocket pairing is thus preferred. 
For Case II, where $\mu_{c,2} > \mu > E^W_{++} = 8.6$meV, there are inversion-related Chern number $C=\pm 1$ electron pockets contained in a larger topologically trivial electron pocket when $\mu_{c,2} > \mu > \mu_{c,1}$, as shown in Fig. \ref{fig:mu002BandFS}. 
For Case III, where $|\mu| < E^W_{++}$, there are two Chern number $C=\pm 2$ Fermi pockets in each half of the Brillouin zone, as shown in Fig. \ref{fig:mu0BandFS} for $k_z > 0$. For the topologically nontrivial pockets in Cases II and III, pairing between inversion-related Fermi pockets is preferred.
The narrow region $\mu_{c,1} > \mu > E^W_{++}$, where there are four $C=\pm 1$ Fermi pockets, may be an artifact of extending the linearized model to the $\Gamma$ point and may not exist in the real material 
\footnote{See Fig. 2c in Ref. \cite{Baidya2020}}. 
As the chemical potential is brought further below $\epsilon_D$, $C = \pm 1$ and trivial $C = 0$ Fermi pockets similar to those in Cases I and II emerge \cite{Baidya2020}.

\begin{figure*}
\subfigure[]{\centering \epsfig{file=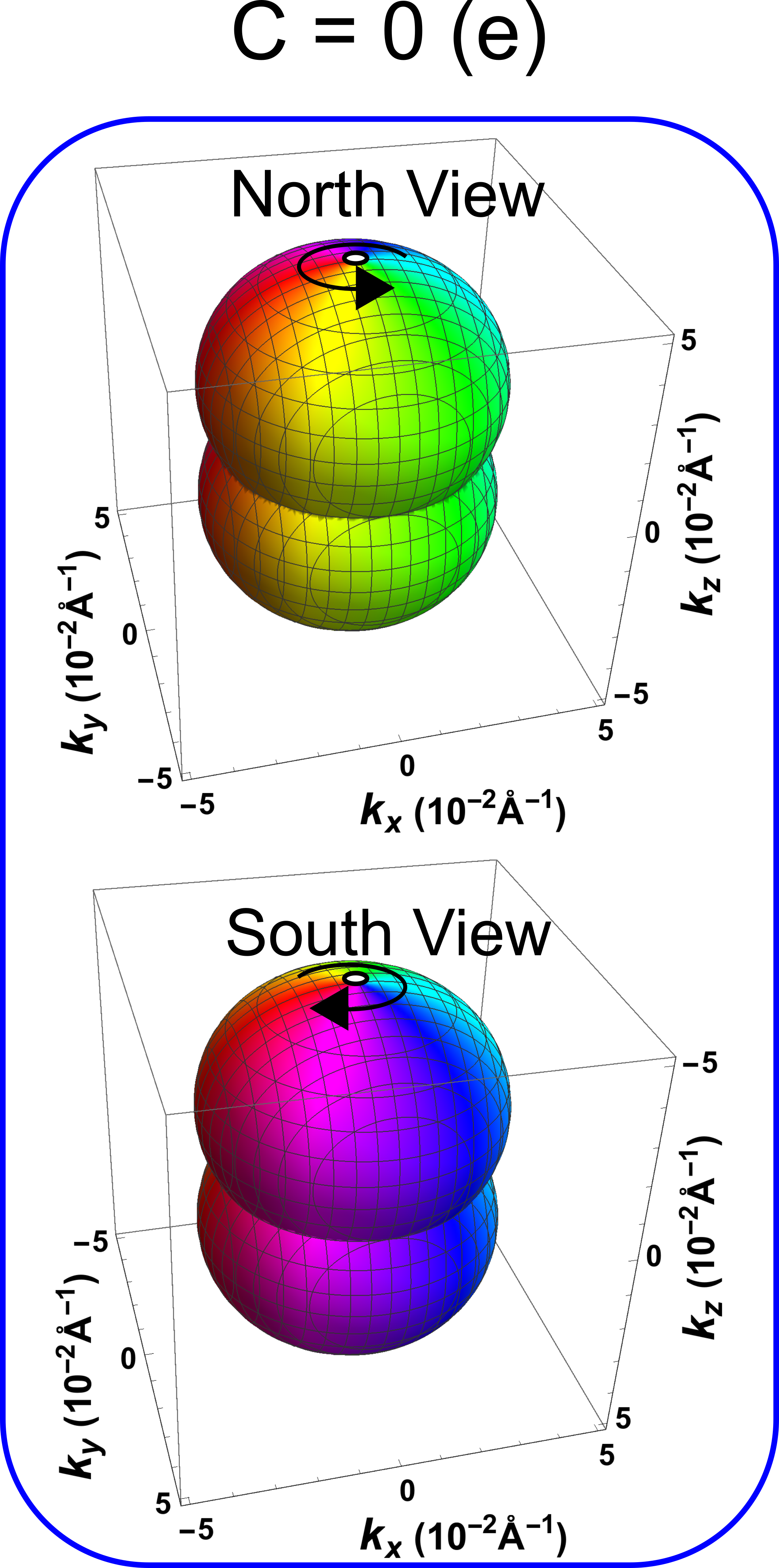, width=0.2\linewidth}
\label{fig:mu0035phase}}
\subfigure[]{\centering \epsfig{file=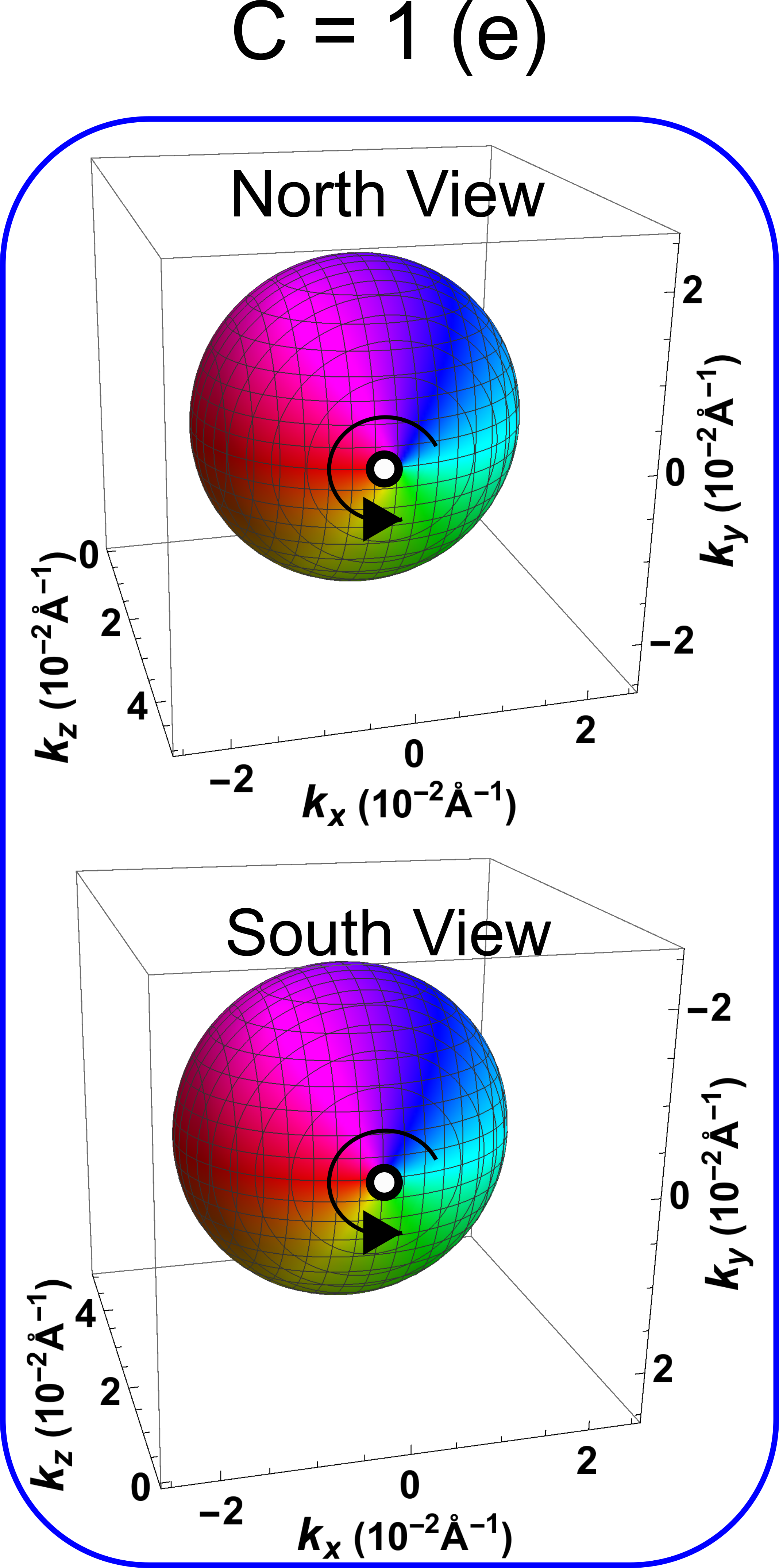, width=0.2\linewidth}
\label{fig:mu002phase}}
\subfigure[]{\centering \epsfig{file=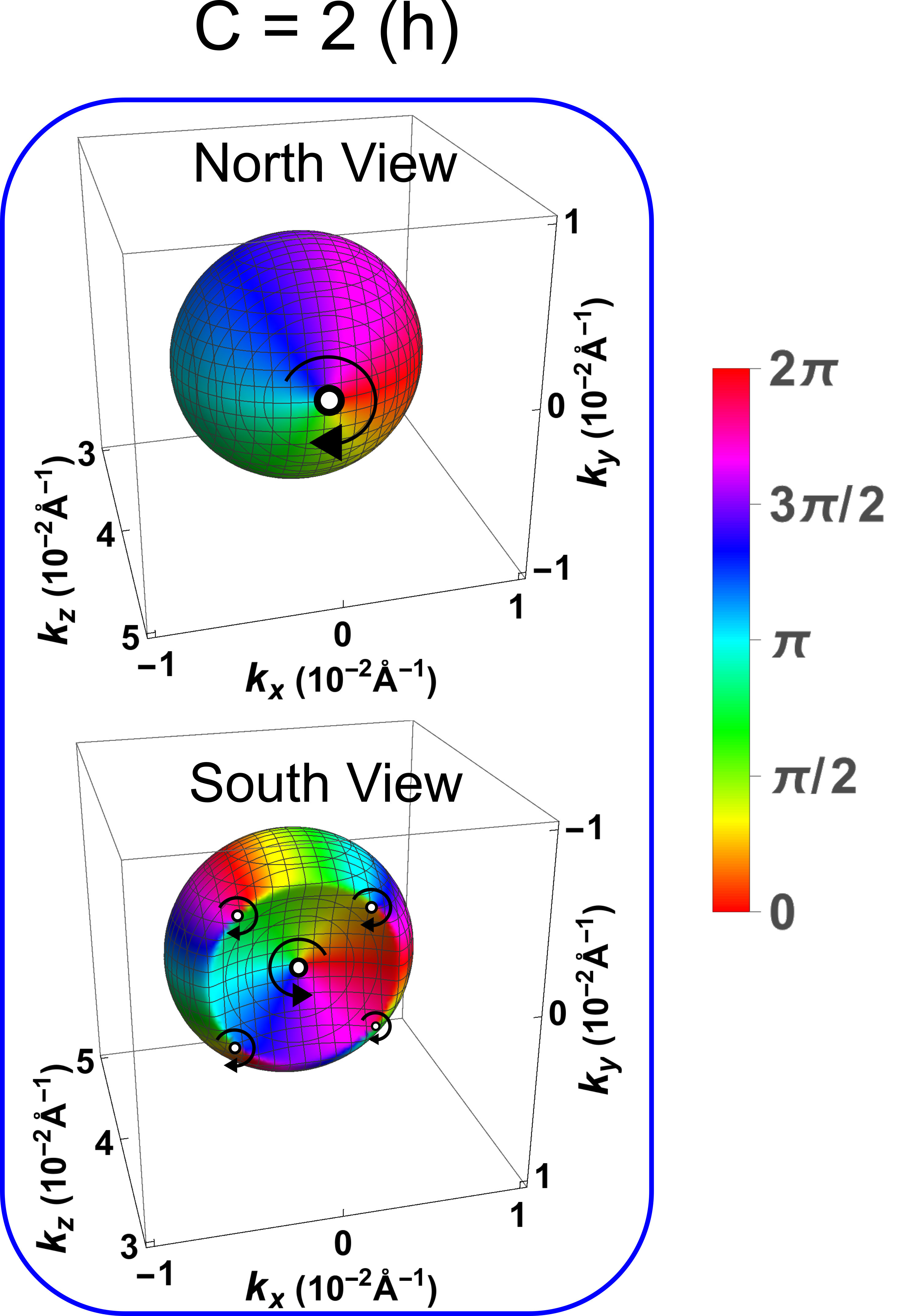, width=0.276\linewidth}
\label{fig:mu0phase}}
\caption{Phase winding patterns of the band projected gap function $\Delta_{proj}$ for pairing matrix $\hat{\Delta}(\k)=\hat{\Delta}_{B_u}$ for ($a$) the smaller electron pocket of $C=0$ in Case I, ($b$) the electron pocket of $C=1$ in Case II and ($c$) the hole pocket of $C=2$ in Case III. Here, $\beta_s = \beta_p= 5.4\times 10^{-5} eV/T$. 
The pairing phase is shown in color, and around each gap node, indicated by a dot,
a circular arrow is drawn along the phase winding direction. 
In ($a$), the north (south) view has the $k_z$ axis pointing up (down), while in ($b$) and ($c$), the north (south) view has the $k_z$ axis pointing into (out of) the page, and the pairing phase is shown in the appropriate gauge for smooth eigenstates at the north (south) pole.
}
\label{fig:g02phase}
\end{figure*}

\textit{Monopole superconductivity with different pair monopole charges.} 
Monopole superconductivity with nonzero pair monopole charge $q_p$ differs from ordinary $s$-, $p$-, or $d$-wave superconductivity by a nonvanishing total momentum-space vorticity $\nu_{tot} = 2q_p$ of the gap function \cite{Murakami2003a,Li2018}. 
For pairing between Fermi pockets with opposite Chern numbers $\pm C$, $q_p = C$, and the pairing order for nonzero pair monopole charge is in the same topological class as the charge-$q_p$ 
monopole harmonics 
\cite{Wu1976}. 
The possible pairing order is then constrained by the total vorticity $\nu_{tot}$ and can be identified as monopole superconductivity when $\nu_{tot} \neq 0$. To characterize the possible pairing orders, we will consider the pairing in an irreducible representation (irrep) of $C_{4h}$, which are all one-dimensional since the point group is abelian. Thus, the pairings we consider satisfy $D(g) \hat{\Delta}(\k) D(g)^T = \chi(g) \hat{\Delta}(g\k)$, with $D(g)$ the matrix representation for $C_{4h}$ group element $g$ in the four-component spin-orbit coupled basis of ${\mathcal{H}}(\k)$ and $\chi$ the character of the representation. We will first discuss the different pairing orders that can arise for the different $q_p$ in Cases I, II, and III for an example pairing in the $B_u$ representation, a simple momentum-independent pairing $\hat{\Delta}_{B_u} = \Delta_0 M_{02}$. 
Here $M_{ij} = \sigma_i \otimes \tau_j$ for the matrix part of the gap function, with $i=0,1,2,3$ labeling the identity and $x$, $y$, $z$ Pauli matrices in spin and orbital space.

To obtain the effective low-energy pairing order analytically,
we consider a simplified model with equal Zeeman couplings for the $s$ and $p$ orbital bands, $\beta_s = \beta_p = 5.4\times 10^{-5} eV/T$. This is a continuous deformation of the band structure that does not introduce additional band crossings or otherwise change the Weyl node structure and thus is topologically equivalent to the $\k \cdot \p$ model. In particular, this deformation slightly changes the shape but not the Chern numbers of the Fermi pockets in Cases I, II, and III, as shown in S. M. Sec. III. 

The vorticity at each gap node, which is gauge invariant, can be identified with the winding of the pairing phase near the node in an appropriate smooth gauge and calculated from the paired states near the Fermi surface.
The momentum-space vorticity at the $i$th gap node is $\nu_i=\oint_{L_i} d\mathbf{l}_{\k} \cdot \mathbf{v}_\k/(2\pi)$, where $\mathbf{v}_{\k} = \nabla_\k \varphi_\k - \mathbf{A}_p(\k)$ is the gauge-invariant circulation field in terms of the pairing phase $\varphi_\k$ and the pair Berry connection $\mathbf{A}_p(\k)$, with integration performed over an infinitesimally small loop $L_i$ surrounding the node that is oriented counterclockwise relative to the normal direction \cite{Murakami2003a,Li2018}. 
When the gauge is chosen such that the pair Berry connection $\mathbf{A}_p(\k)$ is smooth near the $i$th node, the vorticity is the winding number of the pairing phase $\varphi_\k$ of the Fermi surface-projected gap function. 
In the weak-coupling regime, the Fermi surface projection selects the relevant paired states in the band basis, as described by the projected gap function 
\begin{equation}
    \Delta_{proj}(\k) = \langle \psi_\k | \hat{\Delta}(\k) | \psi^h_\k\rangle,
\end{equation}
with $|\psi_\k\rangle$ the band eigenstate of $\mathcal{H}(\k)$ at the Fermi pocket at $\k$ and  $|\psi_\k^h\rangle$ the eigenstate of the hole counterpart $-\mathcal{H}^*(-\k)$ at $-\k$. 
The superconducting phase winding patterns and monopole harmonic structures can then be studied by examining the phase of $\Delta_{proj}(\k)$ over each Fermi pocket with the appropriate smooth gauge near each node. Further details on the relationship between the monopole charge, Chern number, projected gap function, and gauge structure can be found in S. M. Sec II.

We first consider intra-Fermi pocket pairing for a topologically trivial electron pocket.
In Case I, $\mu = 35$meV, where the Fermi surface consists of two topologically trivial electron pockets, the phase winding of the projected gap function is shown over the smaller of the two Fermi pockets with pairing matrix $\hat{\Delta}_{B_u}(\k)$ in Fig. \ref{fig:mu0035phase}. Since the Fermi pocket has Chern number $C=0$, the relevant eigenstate can be smoothly defined with a single choice of gauge. The projected gap function has nodes along the $k_z$ axis with phase winding around the nodes. The vorticity at each node, or the counterclockwise phase winding number relative to the normal direction, is $+1$ at the north pole and $-1$ at the south pole, 
and the total vorticity of the gap function over the Fermi pocket vanishes. These two gap nodes, near which the BdG quasiparticle spectrum is linear, are emergent BdG Weyl nodes with opposite chiralities. Similarly to the smaller Fermi pocket, the larger pocket in this case is topologically trivial, and the corresponding gap function has qualitatively similar behavior.

We next consider the pairing between the topological pockets in Case II, $\mu = 20$meV, where the Fermi surface consists of a large, trivial electron pocket surrounding two smaller, topological electron pockets with $C = \pm 1$ that are related by parity. Consider the smaller pocket with $C = +1$ at $k_z>0$. In this case, different gauge choices (see S. M. Sec. III) are needed near the north and south poles of the pocket to enforce locally smooth eigenstates and identify local phase winding with the vorticity.
The phase of the projected gap function for $\hat{\Delta}_{B_u}(\k)$ is shown near the north and south poles in Fig. \ref{fig:mu002phase}. 
The phase winds counterclockwise with respect to the normal direction around the nodes at both poles, and the total vorticity is thus $+2$, consistent with $q_p = 1$ monopole order due to pairing between the $C=+1$ and $C=-1$ Fermi pockets. 
Like in Case I, the BdG spectrum near the gap nodes is linear. However, unlike in Case I, in Case II, pairing between the $C=+1$ and $C=-1$ pockets leads to emergent BdG Weyl nodes with the same chirality. 

We finally consider Case III, $\mu = 0$meV, where the Fermi surface consists of four Fermi pockets with $C = \pm 2$. We focus on the $C=+2$ hole pocket, the pocket furthest from the $\Gamma$ point in the $+\hat{k}_z$ direction, which participates in inter-Fermi pocket pairing with the $C=-2$ hole pocket at $k_z < 0$. The pocket 
consists of a concave region near the south pole and a convex region in the northern hemisphere. Note that since the vorticity and Chern number are calculated for electron states, the normal direction should be taken to point inwards for this hole pocket. In the appropriate gauges near the two poles, we see in Fig. \ref{fig:mu0phase} vorticity $+1$ near the north pole and vorticity $-1$ near the south pole as well as four additional nodes along the boundary of the concave region each with vorticity $+1$. Thus, the total vorticity, $+4$, is consistent with $q_p = 2$ pairing between $C=+2$ and $C=-2$ pockets. For the pairing matrix considered here, all six nodes are linear, corresponding to emergent BdG Weyl quasiparticles with opposite chiralities near the nodes at the two poles and the same chirality near the remaining four nodes.

In all three cases, we see that the total vorticity is equal to twice the pair monopole charge, $\nu_{tot} = 2q_p = 2C$. Magnetically doped Cd$_3$As$_2$ in principle allows for Fermi pockets with Chern numbers $C=0, 1,$ or $2$, depending on the chemical potential, which in turn feature different gap function structures. Superconducting pairing between $C = 1$ or $2$ Fermi pockets and the parity-related $C = -1$ or $-2$ Fermi pockets leads to a gap function describing $q_p = C \neq 0$ pair monopole charge \cite{Li2018} that can be recognized from the vorticity of the gap function projected to the Fermi surface. Compared to the $C=1$ case, monopole superconductivity involving $C=2$ Fermi pockets requires a larger total vorticity, $+4$ rather than $+2$, and thus necessitates a richer gap function structure that either features more nodes, as in Fig. \ref{fig:g02phase}, or features nodes with different phase winding patterns. 

\begin{figure*}
\subfigure[]{\centering \epsfig{file=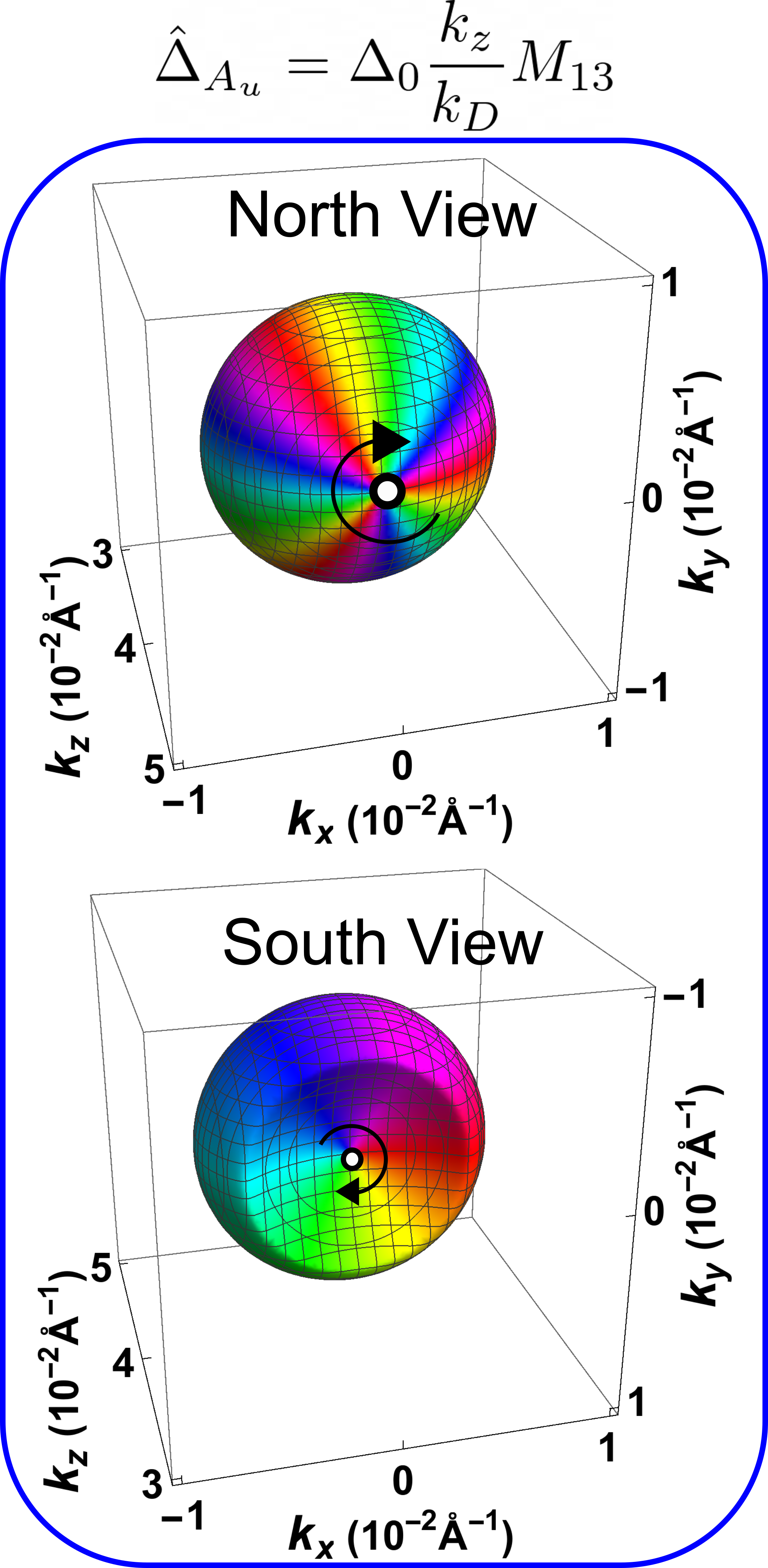, width=0.175\linewidth}
\label{fig:mu0_g13}}
\subfigure[]{\centering \epsfig{file=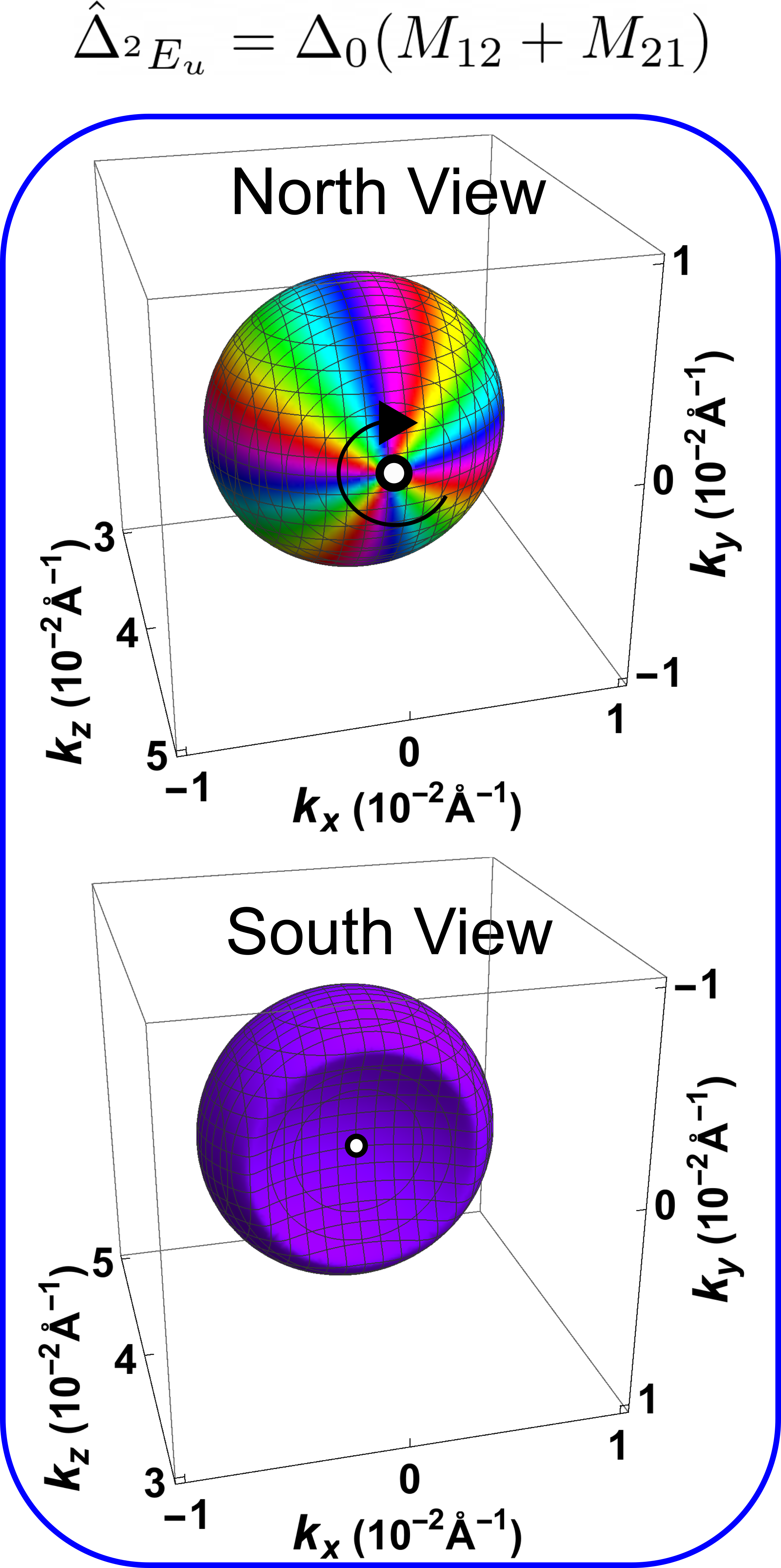, width=0.175\linewidth}
\label{fig:mu0_g12p21}}
\subfigure[]{\centering \epsfig{file=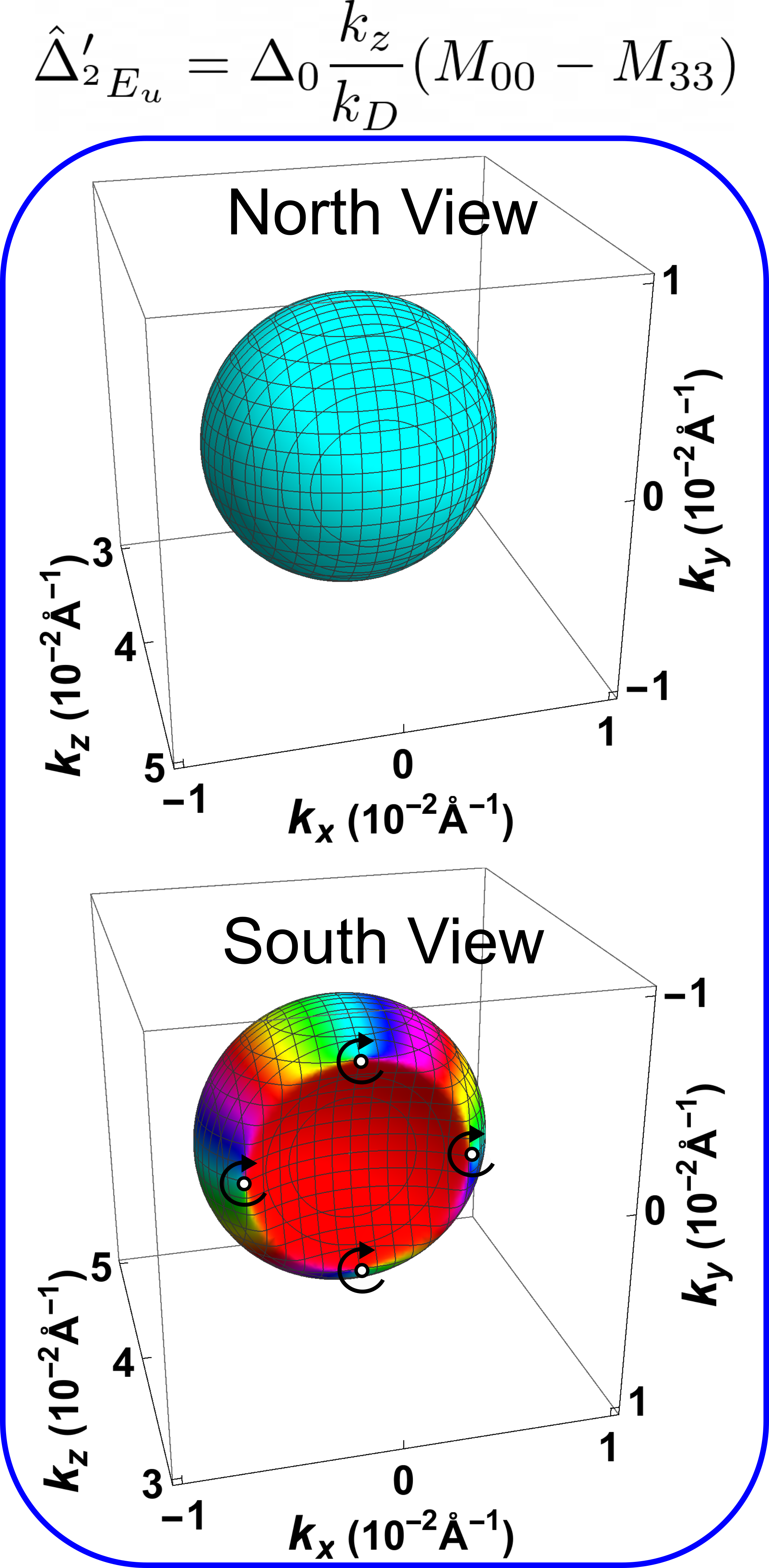, width=0.175\linewidth}
\label{fig:mu0_g00m33}}
\subfigure[]{\centering \epsfig{file=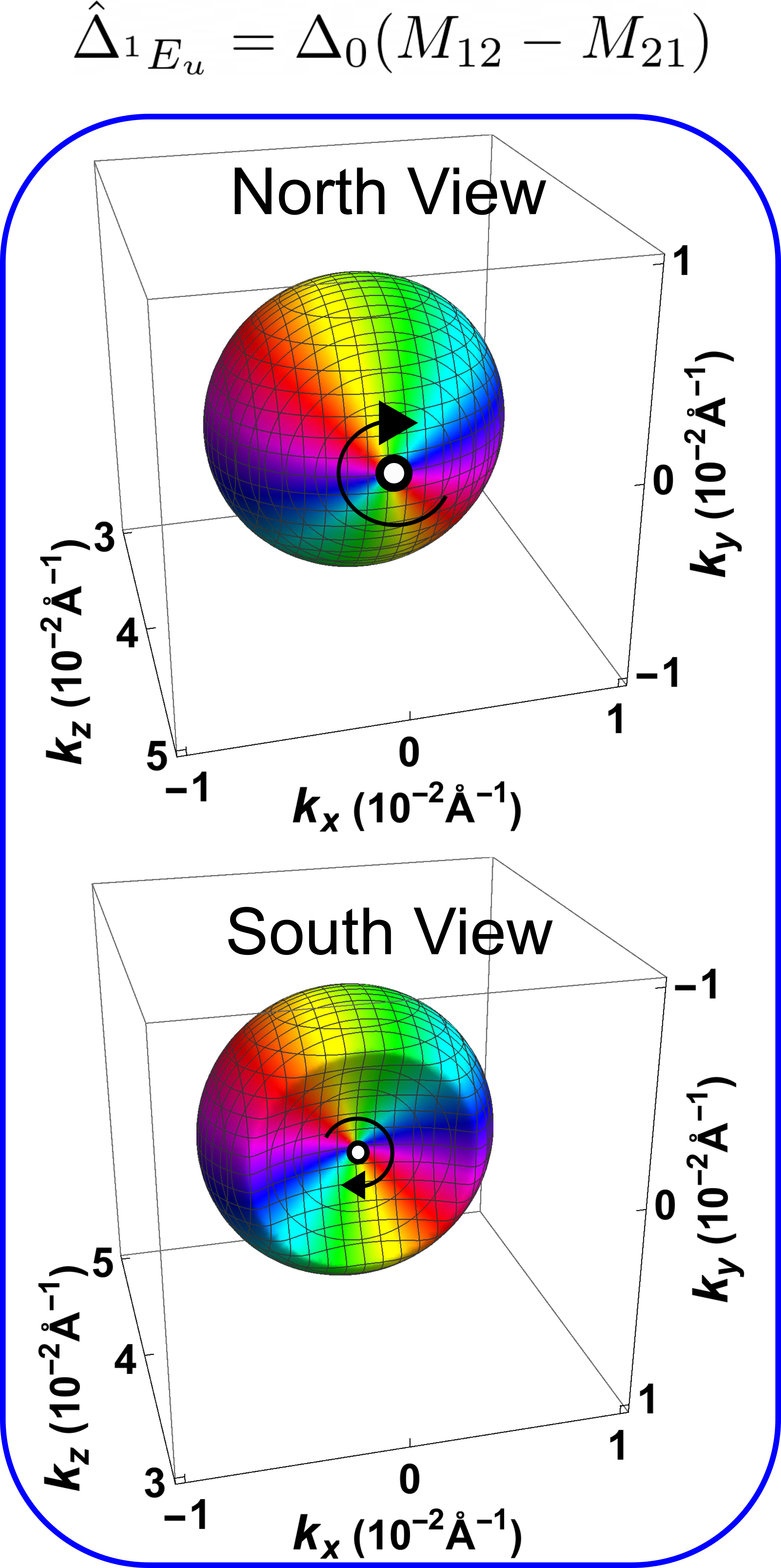, width=0.175\linewidth}
\label{fig:mu0_g12m21}}
\subfigure[]{\centering \epsfig{file=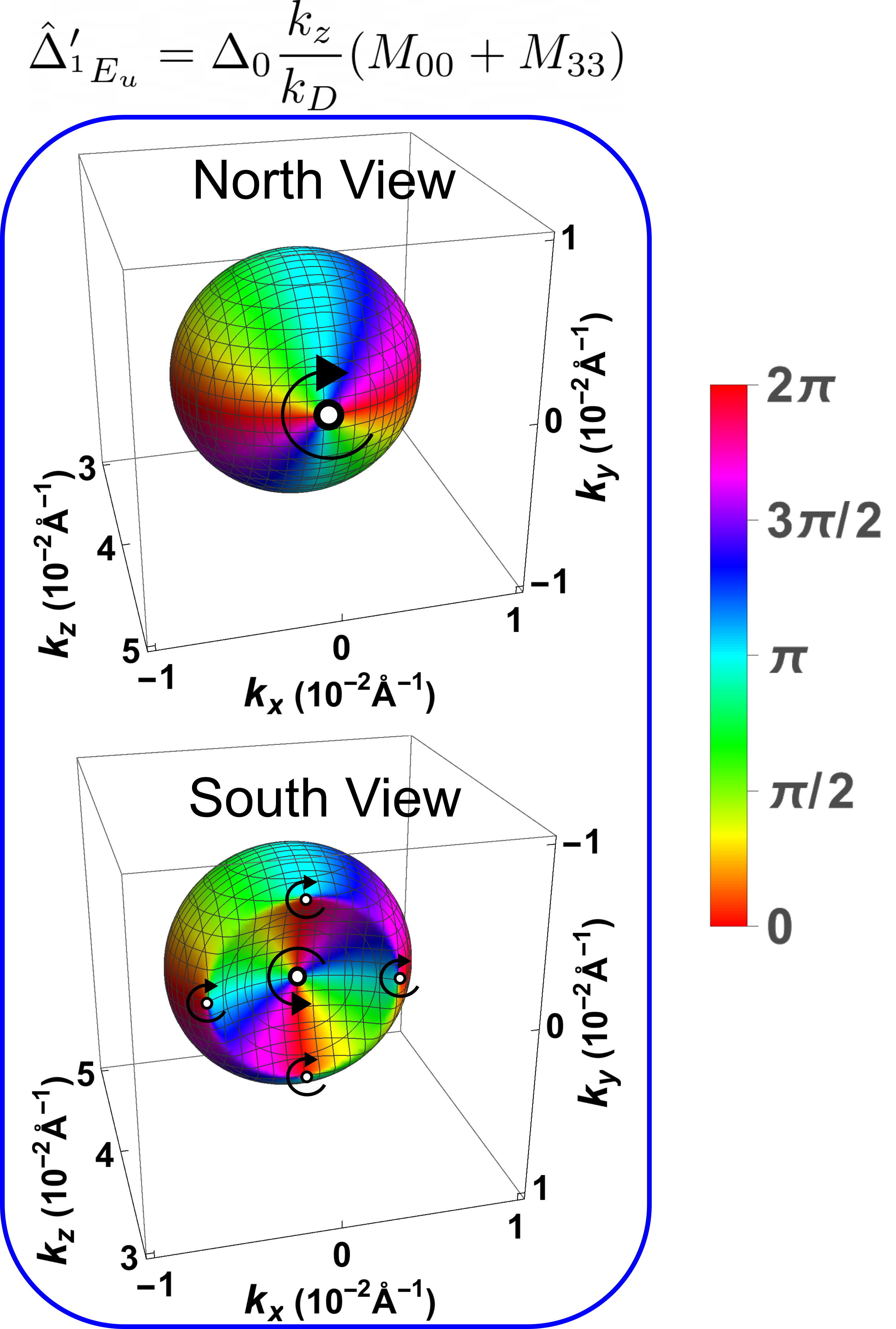, width=0.24\linewidth}
\label{fig:mu0_g00p33}}
\caption{Phase winding patterns of the projected gap function at the $C=2$ hole pocket in Case III with pairing matrices
($a$) $\hat{\Delta}_{A_u}$, 
($b$) $\hat{\Delta}_{^2E_u}$, 
($c$) $\hat{\Delta}_{^2E_u}'$, 
($d$) $\hat{\Delta}_{^1E_u}$ 
and ($e$) $\hat{\Delta}_{^1E_u}'$
that together with the $B_u$ pairing in Fig. \ref{fig:mu0phase} account for all qualitatively distinct patterns. 
The parameters, 
color scheme, and views are the same as those in Fig. \ref{fig:mu0phase}. 
}
\label{fig:IrrepPhase}
\end{figure*}

\textit{Pair monopole charge-2 superconductivity with different symmetries.}
For monopole superconductivity with a fixed pair monopole charge, the crystalline symmetry allows for different pairings that in general lead to different pairing phase winding patterns, subject to the topologically constrained fixed total vorticity over the Fermi pocket. 
In Case III, where richer patterns occur due to the higher pair monopole charge $q_p = 2$, the phase winding patterns from representative pairings in the one-dimensional $A_u$, $^2E_u$, and $^1E_u$ representations with minimal momentum dependence are shown in Fig. \ref{fig:IrrepPhase}. Together with the $B_u$ pairing in Fig. \ref{fig:g02phase}, these pairings account for each of the irreps allowed by symmetry and Fermi statistics. For the $A_u$ pairing in Fig. \ref{fig:mu0_g13}, $\hat{\Delta}_{A_u} = \Delta_0 \frac{k_z}{k_D}M_{13}$, which is fully gapped over the shown Fermi pocket, the projected gap function exhibits nodes along the $k_z$ axis with vorticity $+3$ at the north pole and $+1$ at the south pole of the pocket.
Within the $^2E_u$ and $^1E_u$ representations, there are two qualitatively different phase winding patterns, since there are different pairing matrices corresponding to different bases of the representation. 
The first of the $^2E_u$ pairings, $\hat{\Delta}_{^2E_u}=\Delta_0(M_{12}+M_{21})$ with an antisymmetric pairing matrix, leads to the phase winding pattern shown in Fig. \ref{fig:mu0_g12p21} that features nodes along the $k_z$ axis with phase winding only at the north pole, contributing $+4$ vorticity. In contrast, the second $^2E_u$ pairing, $\hat{\Delta}_{^2E_u}'=\Delta_0\frac{k_z}{k_D}(M_{00}-M_{33})$ with a symmetric pairing matrix, features a projected gap function that is gapped at the north and south poles and instead features four nodes each contributing vorticity $+1$ at azimuthal angles of $\phi_\k = 0, \pm \pi/2,$ and $3\pi/2$ along the boundary of the concave region near the south pole, as shown in Fig. \ref{fig:mu0_g00m33}. The phase winding patterns due to pairing in the $^1E_u$ representation, $\hat{\Delta}_{^1E_u}=\Delta_0(M_{12}-M_{21})$ in Fig. \ref{fig:mu0_g12m21} and $\hat{\Delta}_{^1E_u}'=\Delta_0\frac{k_z}{k_D}(M_{00}+M_{33})$ in Fig. \ref{fig:mu0_g00p33}, both feature nodes with vorticity $+2$ at the north pole, but while the $\hat{\Delta}_{^1E_u}$ pairing features a $+2$ vorticity node at the south pole, the $\hat{\Delta}'_{^1E_u}$ pairing features $-2$ vorticity at the south pole with four additional nodes each with $+1$ vorticity along the boundary of the concave region. Further details about the phase winding patterns, including the patterns that arise for trivial and $C=\pm 1$ Fermi pockets in the $A_u$, $^2E_u$, and $^1E_u$ representations, can be found in S. M. Sec. IV.

\textit{Discussion.} 
In magnetically doped Cd$_3$As$_2$, the topological Fermi pockets allow for the potential realization of monopole superconductivity. 
We saw for example pairing in the $B_u$ representation that
the total winding number of the pairing phase over an entire Chern number $C$ Fermi pocket is an observable topological invariant $\nu_{tot} = 2q_p = 2C$. 
As the chemical potential, and hence the Chern number of the Fermi pocket, varies, the low energy superconducting pairing order goes through topological transitions where the phase windings at certain gap nodes reverse, as in the $C=1$ case compared to the topologically trivial case shown in Figs. \ref{fig:mu002phase} and \ref{fig:mu0035phase} respectively, or where more gap nodes or higher phase winding numbers emerge, as in the $C=2$ case compared to $C=1$ in Figs. \ref{fig:mu0phase} and \ref{fig:mu002phase}.
In each case, the specific gap function structure can differ for different pairings but is constrained by the total phase winding.
Varying the pairing itself for fixed $C$ similarly allows for different phase winding patterns subject to the fixed $\nu_{tot}$ constraint, as in the example pairings from each remaining allowed irrep of the $C_{4h}$ point group shown in Fig. \ref{fig:IrrepPhase}. Pairings in the same topological class with distinct nodal structures, such as the $A_u$ and $B_u$ pairings of Figs. \ref{fig:mu0_g13} and \ref{fig:mu0phase}, can potentially be distinguished by heat capacity measurements. Further information about the pairings in the different irreps can be found in S. M. Sec. IV, including phase winding patterns for $C=0$ and $C=1$ Fermi pockets. 

For monopole superconductivity with pair charge $q_p$, the gap function is characterized by $l \geq q_p$ angular momentum channels, a characteristic of monopole harmonic symmetry.
As a result, an interesting direction for future work is to consider a suitable Weyl semimetal with six-fold rotational symmetry protecting Weyl nodes with topological charge $3$ \cite{Fang2012}. 
This type of system could potentially be realized by magnetically doping a Dirac semi-metal with cubic Dirac nodes, such as the candidate materials $A$ Mo$X_3$ ($A=$ Na, K, Rb, In, Tl; $X=$ S, Se, Te] \cite{Liu2017, Lv2021}.
Pairing between Fermi pockets with Chern numbers $\pm 3$ in such a system would in principle lead to the monopole analog of $f$-wave pairing as the lowest allowed partial wave channel.

\textit{Acknowledgements.}
We thank David Vanderbilt for the helpful discussion on the Chern numbers of Fermi surfaces in magnetically doped Cd$_3$As$_2$. We acknowledgement the Institute for Quantum Matter, an Energy Frontier Research Center funded by the U.S. Department of Energy, Office of Science, Basic Energy Sciences under Award No. DE-SC0019331, for support on building the appropriate model and calculating the initial examples of monopole pairing in this material. We also acknowledge the NSF CAREER Grant DMR-1848349 and the Alfred P. Sloan Research Fellowships under grant FG-2018-10971, which supported the completion of our discussion by scanning and sorting possible pairings according to all irreducible representations in this system.

%

\end{document}


\title{Supplemental Material for ``Monopole Superconductivity in Magnetically Doped Cd$_3$As$_2$"}
\maketitle

\tableofcontents

\section{$\k\cdot\p$ models of C\lowercase{d}$_3$A\lowercase{s}$_2$}

\subsection{Symmetry analysis for Dirac semimetal C\lowercase{d}$_3$A\lowercase{s}$_2$}
\label{app:symm}

We review the form of the $\k\cdot\p$ Hamiltonian for the Dirac semimetal Cd$_3$As$_2$ \cite{Wang2013}. The $D_{4h}$ point group together with time reversal symmetry of the material in the absence of magnetic dopants restricts the structure of the $\k\cdot\p$ Hamiltonian \cite{Zhang2010b}. Following the theory of invariants \cite{Winkler}, the symmetry condition $D(g) \mathcal{H}(g^{-1}\k) D(g)^{-1} = \mathcal{H}(\k)$ implies that for a particular basis of matrices, the Hamiltonian matrix is a sum of matrices paired with functions of $\k$ that both belong to the same irreducible representation of $D_{4h}$. 

As discussed in Ref. \cite{Wang2013}, the bands relevant to the Dirac crossing constitute the spin-orbit coupled basis $\{|S_{J=\frac{1}{2}}, J_z=\frac{1}{2}\rangle, |P_{\frac{3}{2}}, \frac{3}{2}\rangle, |S_{\frac{1}{2}}, -\frac{1}{2}\rangle, |P_{\frac{3}{2}}, -\frac{3}{2}\rangle\}^T$. The symmetry group is generated by time reversal $\Theta$, inversion $\Pi$, rotation $C_4$ about the $z$ axis, and rotation $C_2'$ about the $x$ axis, which in the spin-orbit coupled basis can be written as
\begin{equation}
    \begin{aligned}
    D(\Theta) &= i\sigma_y\otimes I \circ \mathcal{K}\\
    D(\Pi) &= I \otimes \tau_z \\
    D(C_4) &= e^{-i\frac{\pi}{4} \Sigma} \equiv e^{-i\frac{\pi}{4}\sigma_z\otimes(2I - \tau_z)}\\
    D(C_2') &= -i\sigma_x \otimes \tau_z,
    \end{aligned}
\end{equation}
with the $\tau_i$ Pauli matrices acting on the orbital part of the basis (s/p) and $\sigma_i$ acting on the spin part, with spin $\uparrow/\downarrow$ corresponding to $\mathrm{sgn}(J_z) = \pm$. Here $\mathcal{K}$ is the complex conjugation operator. The full double group need not be considered since $2\pi$ spin rotation acts as $-I\otimes I$ and thus commutes with $\mathcal{H}(\k)$. The presence of both time reversal and parity requires that the gamma matrices appearing in $\mathcal{H}(\k)$ be even under the product operation $\Theta \Pi$, since $\k$ is even under this combined symmetry. This restricts the basis of allowed matrices to $\Gamma_i$ for $i = 0,\dots, 5$ with $\Gamma_0$ the identity, $\Gamma_{1, 2, 3} = \sigma_{x, y, z} \otimes \tau_x$, and $\Gamma_{4, 5} = I \otimes \tau_{y, z}$.

The characters of $\Gamma_i$ under the symmetry transformations $D(\Theta), D(\Pi),$ or $D(C_2')$ are listed in Table \ref{tab:reps}. $\Gamma_1, \Gamma_2,$ and $\Gamma_5$ form one-dimensional representations while $(\Gamma_3, \Gamma_4)$ is an irreducible two-dimensional representation, since $C_2'$ and $C_4$ respectively act as $\sigma_z$ and $-i\sigma_y$ on $(\Gamma_3, \Gamma_4)$ and thus cannot be simultaneously diagonalized. As a result, comparison to the character table \cite{Altmann1994} for $D_{4h}$ uniquely identifies the corresponding irreducible representations, listed in Table \ref{tab:reps} along with the corresponding basis functions to lowest nonzero order in $\k$.

\begin{table}[ht]
 \begin{tabular}{c|c c c c | c | c}
     $\Gamma_i$ & $\Theta$ & $\Pi$ & $C_2'$ & $C_4$ & Rep & Functions\\
     \hline\hline
     $\Gamma_0 = I\otimes I$ & $1$ & $1$ & $1$ & $1$ & $A_{1g}$ & $k_x^2 + k_y^2, k_z^2$\\
     \hline
     $\Gamma_1 = \sigma_x \otimes \tau_x$ & $-1$ & $-1$ & $-1$ & $-1$ & $B_{2u}$ & $k_z(k_x^2 - k_y^2)$ \\
     \hline
     $\Gamma_2 = \sigma_y \otimes \tau_x$ & $-1$ & $-1$ & $1$ & $-1$ & $B_{1u}$ & $k_x k_y k_z$\\ 
     \hline 
     $\Gamma_3 = \sigma_z \otimes \tau_x$ &&&&& \\ 
     $\Gamma_4 = I \otimes \tau_y$ & $-2$ & $-2$ & $0$ & $0$ & $E_{u}$ & $(k_x, k_y)$\\
     \hline 
     $\Gamma_5 = I \otimes \tau_z$ & $1$ & $1$ & $1$ & $1$ & $A_{1g}$ & $k_x^2 + k_y^2, k_z^2$
  \end{tabular}
\caption{\label{tab:reps} Transformation of $\Gamma_i$ under the $D_{4h}$ symmetries. Comparison to a $D_{4h}$ character table allows the irreducible representations and corresponding basis functions to be identified.}
\end{table}

The symmetry-allowed $\mathcal{H}(\k)$ is then
 \begin{equation}
   \begin{aligned}
     \mathcal{H}(\k) = &\left(e_0 + e_1 k_z^2 + e_2(k_x^2 + k_y^2)\right)\Gamma_0 \\
     &+ \left(M_0 + M_1 k_z^2 + M_2(k_x^2 + k_y^2)\right)\Gamma_5 \\
     &+ A (k_x \Gamma_3 - k_y \Gamma_4) \\
     &+ B_1  k_z(k_x^2 - k_y^2) \Gamma_1 + B_2 2 k_z k_x k_y \Gamma_2,
   \end{aligned}
   \label{eq:Hsymm}
 \end{equation}
 with all coefficients real due to time reversal symmetry and the factor of $2$ on the $B_2$ term added for convenience. The appropriate basis vector in the two-dimensional $E_u$ subspace is selected by requiring $D(g) H(g^{-1}\k) D(g)^{-1} = H(\k)$. The off-diagonal terms in Eq. \eqref{eq:Hsymm} are
 \begin{equation}
     \mathcal{H}_{A,B}(\k) = \pmat{0 & A k_+ & 0 & f_B(\k)\\
     A k_- & 0 & f_B(\k) & 0\\
     0 & f_B^*(\k) & 0 & -A k_- \\
     f_B^*(\k)  & 0 & -A k_+ & 0}
 \end{equation}
 with $f_B(\k) \equiv G k_z k_-^2 + G 'k_z k_+^2$ and $G, G' = \frac{B_1 \pm B_2}{2}$ for $k_{\pm} = k_x \pm i k_y$. In particular, $f_B(\k)$ includes only the $k_-^2$ term when $B_1 = B_2$, which is not generally required by the $D_{4h}$ symmetry. 
 
 As noted in Ref. \cite{Baidya2020}, the $G'$ term arises due to weak crystal field effects and can be taken to be zero. In fact, the inclusion of a small $G' < G$ does not affect topological properties of the magnetically doped model, such as the Chern number and hence the number of BdG nodes required by the topological pairing. 

\subsection{Model parameters for magnetically doped C\lowercase{d}$_3$A\lowercase{s}$_2$}

In the presence of magnetic dopants, modeled by a Zeeman field along the $z$ direction, the Weyl points near $k_z = k_D$, the location of the Dirac node in the absence of magnetic dopants, are described by
\begin{equation}
\begin{aligned}
   &{\cal{H}}(\k) = (\epsilon_D - \mu) I +
    \begin{pmatrix}
        v_s\tilde{k}_z + \beta_s h & A k_+ & 0 & G k_-^2 \\
        A k_- & -v_p\tilde{k}_z + \beta_p h & G k_-^2 & 0 \\
        0 & G k_+^2 & v_s\tilde{k}_z - \beta_s h & -A k_- \\
        G k_+^2 & 0 & -A k_+ & -v_p \tilde{k}_z - \beta_p h
    \end{pmatrix},
    \label{eq:H}
\end{aligned}
\end{equation}
as studied in Ref. \cite{Baidya2020}, with fitting parameters reproduced in Table \ref{tab:params} for convenience.

\begin{table}[h]
\begin{tabular}{c|c|c|c}
    $v_s$(eV $\si{\mathring{A}}$) & $v_p$(eV $\si{\mathring{A}}$) & $A$ (eV $\si{\mathring{A}}$) & $G$ (eV $\si{\mathring{A}^2}$)\\
    \hline
    $2.68$ & $0.56$ & $0.99$ & $10$ \\
    \hline \hline
    $\beta_s$ (eV/T) & $\beta_p$ (eV/T) & $h$ (T) & $k_D$ ($\si{\mathring{A}^{-1}}$)\\
    \hline
    $5.4\times 10^{-5}$ & $1.15 \times 10^{-4}$ & $100$ & $0.037$

\end{tabular}
\caption{\label{tab:params} Parameters in the $\k\cdot\p$ Hamiltonian Eq. \eqref{eq:H}, as calculated in Ref. \cite{Baidya2020}.}
\end{table}

 \section{Monopole charge and Chern number}

In this section, we review the relationship between monopole charge and Chern number associated with topological Bloch states and the corresponding monopole pairing order. 
In particular, we review how an obstruction to smoothly defining the gauge for Bloch states over a Fermi surface indicates a nonzero Chern number 
as well as a nonzero monopole charge associated with the corresponding pairing order. 

\subsection{Monopole charge and Chern number for Bloch states}
\label{app:Chern}

The topological structure of a Fermi surface with a nonzero Chern number closely resembles that of a Dirac magnetic monopole.  
Consider a Bloch state  $|\psi_\k \rangle$ on a spherical Fermi surface. 
For a Fermi surface with Chern number $C$, the Bloch state at the Fermi surface experiences a fictitious magnetic field $\mathbf{B}_\k=q \hat{\mathbf{k}}$ from a momentum-space monopole charge $q \equiv \frac{1}{4\pi}\iint_{FS} \nabla \times \mathbf{A}_\k \equiv C/2$. 
The corresponding vector potential, the Berry connection $\mathbf{A}_\k=i\langle \psi_\k|\mathbf{\nabla}_\k|\psi_\k\rangle$, is piecewise well defined over the northern and the southern hemispheres. 

The Bloch states $\psi_\k$ cannot be smoothly defined globally over the entire Fermi surface when $q\neq 0$ but can be smoothly defined locally over the north and the south hemispheres under appropriate gauges related at the equator by
\begin{equation}
  |\psi^N\rangle = e^{-2i q \phi_\k} |\psi^S\rangle.
  \label{eq:transition}
 \end{equation}
 Here $N$ and $S$ refer to the gauges for which the states are smooth over the patches $\theta_\k < \pi/2 + \delta$ covering the northern hemisphere and $\theta_\k > \pi/2 - \delta$ covering the southern hemisphere for a small angle $\delta > 0$. Eq. \eqref{eq:transition} describing the transition function then holds on the overlap, $\pi/2 + \delta \geq \theta_\k \geq \pi/2 - \delta$.
 The components of $|\psi\rangle$ can be understood as sections of a complex line bundle and written in terms of monopole harmonics \cite{Wu1976}.

 
 The Chern number can be calculated from the Berry connection $A^j = \langle \psi^j|i\nabla_\k|\psi^j\rangle$ in each patch $j=N,S$. From the transition function, $A^N = A^S + \frac{2q}{\sin\theta_\k}\hat{\phi}$. The Chern number is then
 \begin{equation}
 \begin{aligned}
 C &\equiv \frac{1}{2\pi} \left[\iint_{\mathcal{S}^N} \nabla \times A^N + \iint_{\mathcal{S}^S} \nabla\times A^S\right] \\
 &= \frac{1}{2\pi} \oint_{\mathcal{C}_E} A^N - A^S 
  = 2q,
\label{eq:chern}
 \end{aligned}
 \end{equation}
 where $\mathcal{S}^N = \{(\theta_\k, \phi_\k) | \theta_\k\le \pi/2, \phi_\k \in [0, 2\pi)\}$ and $\mathcal{S}^S = \{(\theta_\k, \phi_\k) | \theta_\k\ge \pi/2, \phi_\k \in [0, 2\pi)\}$ are the northern and southern hemispheres and $\mathcal{C}_E = \{(\theta_\k, \phi_\k) | \theta_\k = \pi/2, \phi_\k \in [0, 2\pi)\}$ is the equator oriented in the $+\hat{\phi}$ direction so that the boundary of $\mathcal{S}^N$ is $\mathcal{C}_E$ and the boundary of $\mathcal{S}^S$ is $-\mathcal{C}_E$. Thus, the Chern number can be read immediately from the transition function, and a nontrivial transition function implies a nonzero Chern number. 
 It is important to note that the above Chern number calculation is for an electron pocket, where the normal direction points out of the sphere. For a hole pocket, with normal direction pointing into the sphere, the Chern number for the same transition function is $-2q$.
 
 \subsection{Projected gap function and pair monopole charge}
 
 The projected gap function follows from the paired band eigenstates at the Fermi surface and the form of the pairing matrix. For a pairing matrix $\hat{\Delta}$, the BdG matrix for zero center-of-mass momentum pairing in a parity-symmetric system takes the form
 \begin{equation}
     \mathcal{H}_{BdG} = \pmat{\mathcal{H}(\k) & \hat{\Delta} \\ \hat{\Delta}^\dagger & -\mathcal{H}^*(-\k)},
 \end{equation}
 with the $n$-band system described in a spin-orbital or pseudospin basis by an $n\times n$ matrix $\mathcal{H}$. The particle, $\mathcal{H}(\k)$, and hole, $-\mathcal{H}^*(-\k)$, blocks can be diagonalized by unitary matrices $U_\k$ and $U_{-\k}^*$, where the $j$th column of $U_\k$ is the $j$th eigenstate of $\mathcal{H}(\k)$. Importantly, since eigenstates describing Fermi surfaces with nonzero Chern number require two gauges on two patches covering the Fermi surface to be smoothly defined, the unitary matrix will generally require two gauges as well, related by $U^N_\k = U^S_\k D^{SN}$ on the patch overlap. The $j$th diagonal element of the diagonal matrix $D^{SN}$ is the transition function $e^{-2i q_j \phi_\k}$ for eigenstate $j$. Using $U^{N/S}_{BdG, \k} = \mathrm{diag}[U^{N/S}_\k, U^{N/S *}_{-\k}]$ to diagonalize the band and hole blocks,
 \begin{equation}
     U^{N/S\dagger}_{BdG, \k} \mathcal{H}_{BdG} U^{N/S}_{BdG, \k}
     = \pmat{\mathcal{E}(\k) & U^{N/S\dagger}_{\k} \hat{\Delta} U^{N/S *}_{-\k} \\ \left(U^{N/S\dagger}_{\k} \hat{\Delta} U^{N/S *}_{-\k}\right)^\dagger & -\mathcal{E}(-\k)},
 \end{equation}
 where $\mathcal{E}(\k)$ is an $n\times n$ diagonal matrix with $j$th diagonal element the energy $E_j(\k)$ of the $j$th eigenstate of $\mathcal{H}(\k)$. In this band-diagonal basis, the Fermi surface projection identifies the pairing between states at the Fermi surface as the projected gap function. For a Fermi surface described by eigenstate $j$, $E_j(\k) = 0$, and the projected gap function to lowest order is the component $\left(U^{N/S\dagger}_{\k} \hat{\Delta} U^{N/S *}_{-\k}\right)_{jj}$. Noting that the $j$th column of $U^{N/S}_{\k}$ is the eigenvector $|\psi^{N/S}_j(\k)\rangle$ of $\mathcal{H}(\k)$ and the $j$th column of $U^{N/S *}_{-\k}$ is the eigenvector $|\psi^{N/S}_{h,j}(\k)\rangle$ of the hole block $-\mathcal{H}^*(-\k)$, the projected gap function is
 \begin{equation}
 \begin{aligned}
     \Delta_{\mathrm{proj}}^{N/S} &= \langle \psi^{N/S}_{j}(\k) | \hat{\Delta}| \psi^{N/S}_{h, j}(\k)\rangle \\
     & = \langle \psi^{N/S}_j(\k) | \hat{\Delta} \Pi \mathcal{K}|\psi^{N/S}_j(\k)\rangle.
\end{aligned}
 \end{equation}
 Here $\mathcal{K}$ is the complex conjugation operator and $\Pi$ is the appropriate representation of the parity operator in the spin-orbital or pseudospin basis. Thus, assuming parity symmetry, we write $| \psi^{N/S}_{h, j}(\k)\rangle = \mathcal{K}|\psi^{N/S}_j(-\k)\rangle = \Pi \mathcal{K}|\psi^{N/S}_j(\k)\rangle$. The projected gap function then inherits a transition function from the single-particle eigenstates,
 \begin{equation}
     \Delta_{\mathrm{proj}}^N = e^{4iq_j} \Delta_{\mathrm{proj}}^S. 
 \end{equation}
 Thus, the projected gap function is described by monopole harmonics $Y_{qlm}$ \cite{Wu1976} with $q = 2q_j = C_j$.
 
 Similarly to how the Chern number can be identified from the transition function of a single-particle state, the pair monopole charge can be identified from the transition function of the projected gap function. The pair monopole charge $q_p$ follows from integrating the pair Berry curvature over the Fermi surface and is related to the total vorticity $\nu_{tot} = 2q_p$ \cite{Murakami2003a,Li2018}. As the total vorticity is defined in terms of the integral of the circulation field $\mathbf{v}_\k = \nabla_\k \varphi_\k - A_p(\k)$, the gauge invariance of the circulation field, and hence the vorticity, follows from the pair Berry connection $A_p(\k)$ and pairing phase $\varphi_\k$ transforming the same way. Explicitly, in terms of the projected gap function, the pairing phase is the complex argument $\varphi_\k = \mathrm{arg}[\Delta_{\mathrm{proj}}(\k)]$, and $\nabla_\k \varphi_\k^N = \nabla_\k \varphi_\k^S + \frac{4 q_j}{k \sin \theta_\k} \hat{\phi}_\k$. The pair state transforms with a transition function given by the product of the single-particle transition functions, and thus $A_p^N(\k) = A_p^S(\k)+ \frac{4 q_j}{k \sin \theta_\k} \hat{\phi}_\k$. Integrating the Berry connection $\nabla \times A_p$ over the Fermi surface, a calculation similar to Eq. \eqref{eq:chern} shows $q_p = 2 q_j$, and the pair monopole charge can thus be read from the transition function of $\Delta_{\mathrm{proj}}(\k)$.

\section{Analytical derivation of Chern numbers of Fermi surfaces in a model of magnetically doped C\lowercase{d}$_3$A\lowercase{s}$_2$ with $\beta_s=\beta_p$}
\label{app:vecs}

We consider a deformation of the model in Eq. \eqref{eq:H} with $\beta_s = \beta_p = 5.4\times 10^{-5} eV/T$. This simplification allows the eigenstates to be written in a simpler form where the appropriate gauges for smooth eigenstates near the poles of each Fermi pocket be can be easily identified. This model is a continuous deformation of Eq. \eqref{eq:H} that preserves the topological structure of the Fermi pockets.

The Hamiltonian matrix kernel for $k_z > 0$ can be written as
\begin{equation}
\begin{aligned}
 \mathcal{H}_+(\k) &= \left[v_-(k_z - k_D)-\mu\right] I + \beta_- h \sigma_z \otimes \tau_z + \beta_+h \sigma_z \otimes I \\
 &+ v_+(k_z - k_D) I \otimes \tau_z - A k_y I \otimes \tau_y + A k_x \sigma_z\otimes \tau_x \\
 &+ G(k_x^2 - k_y^2) \sigma_x \otimes \tau_x + G 2k_xk_y \sigma_y\otimes\tau_x \\
 &\equiv f_0(k_z) I + B \Gamma_{12} + \beta_-h \Gamma_{34}+ \sum_{i=1}^5 d_i \Gamma_i
\end{aligned}
\label{eq:HB}
\end{equation}
where $v_\pm = \frac{v_s \pm v_p}{2}$, $\beta_\pm = \frac{\beta_s \pm \beta_p}{2}$, and $B = \beta_+ h$. With $\beta_s \neq \beta_p$, this model is exactly the $\k \cdot \p$ model in Eq. \eqref{eq:H}, here labeled $\mathcal{H}_+$ to emphasize that this linearized model is valid for $k_z > 0$. 
The five anticommuting gamma matrices are $\Gamma_{1,2,3} = \sigma_{x, y, z}\otimes \tau_x$ and $\Gamma_{4, 5} = I\otimes \tau_{y, z}$, as in Sec. \ref{app:symm}, and we denote products of gamma matrices by $\Gamma_{nm} \equiv -i \Gamma_n \Gamma_m$. Thus, $f_0(k_z) = v_-(k_z - k_D) -\mu$, $d_1 = G (k_x^2 -k_y^2)$, $d_2 = G 2k_x k_y$,  $d_3 = A k_x$, $d_4 = -A k_y$, and $d_5 = v_+(k_z - k_D)$.
For the simplified model with $\beta_s = \beta_p$, it involves only six gamma matrices, as $\beta_- = 0$. With $\beta_- = 0$, the energies are
\begin{equation}
 E_{s_1, s_2} = f_0(k_z) + s_1\sqrt{B^2 + d^2 + s_2 2 B \dd},
\end{equation}
labeled by the signs $s_1, s_2 = \pm 1$, and with $\dd \equiv \sqrt{d_3^2 +d_4^2 + d_5^2}$ and $d \equiv \sqrt{d_1^2+d_2^2+(d^{(3)})^2}$.

For $k_z < 0$, the parity-related Hamiltonian is $\mathcal{H}_-(\k) = \Pi \mathcal{H}_+(-\k) \Pi$, with $\Pi = \Gamma_5 = I \otimes \tau_z$. In other words, we consider the model in Eq. \eqref{eq:H} with $\beta_s = \beta_p$ and extended to all $\k$ by requiring parity symmetry,

\begin{equation}
    \mathcal{H}(\k) = 
    \begin{cases}
    \mathcal{H}_+(\k) & k_z > 0 \\
    \mathcal{H}_-(\k) = \Pi \mathcal{H}_+(-\k) \Pi & k_z < 0.
    \end{cases}
\end{equation}
The BdG Hamiltonian, for $k_z > 0$, is thus
\begin{equation}
 \mathcal{H}_{BdG}(\k) = \pmat{\mathcal{H}(\k) & \hat{\Delta}(\k) \\ \hat{\Delta}(\k)^\dagger & -\mathcal{H}^*(-\k)} = \pmat{\mathcal{H}_+(\k) & \hat{\Delta}(\k) \\ \hat{\Delta}(\k)^\dagger & -\Pi \mathcal{H}_+^*(\k)\Pi}, 
\end{equation}
where, in terms of $d_i$, the hole block is
\begin{equation}
\begin{aligned}
\mathcal{H}_h(\k) \equiv -\Pi \mathcal{H}_+^*(\k)\Pi = &-f_0(k_z) -B \Gamma_{12} \\
&+(d_1, -d_2, d_3, -d_4, -d_5)\cdot\vec{\Gamma}.
\end{aligned}
\end{equation}

\begin{figure}[tbp]
\epsfig{file=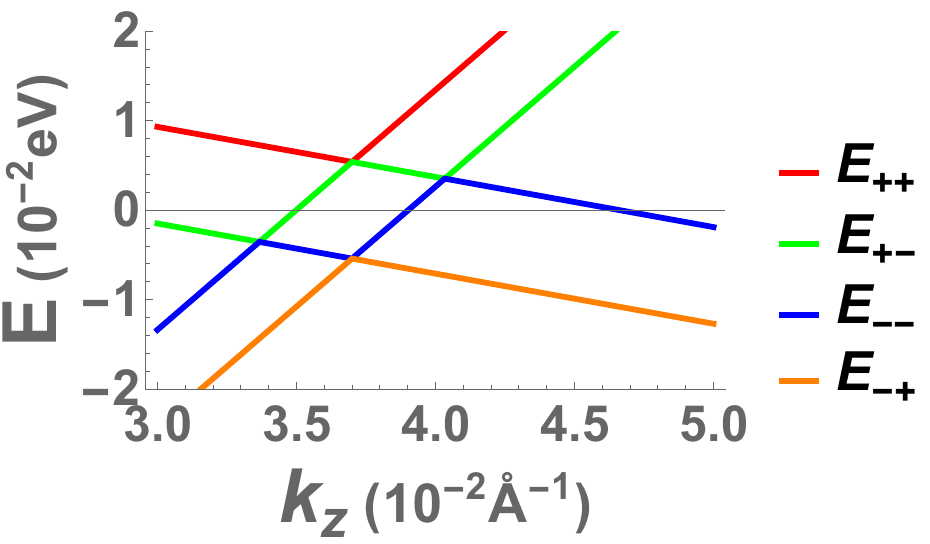, width=0.48\linewidth}
\caption{Spectrum along the $k_z$ axis for $\beta_s = \beta_p$.}
\label{fig:Epmpm}
\end{figure}

The eigenvectors of Eq. \eqref{eq:HB} are, up to an overall phase choice, 
\begin{equation}
\begin{aligned}
 |\psi_{s_1, s_2}\rangle &= \frac{1}{\mathcal{N}}\pmat{ s_2\frac{(d_1 - i d_2)}{d_{12}}d_{34}g^{(3)}_{s_1, s_2}\\ 
\frac{i(d_4-id_3)(d_1 - i d_2)}{d_{34}d_{12}}\left[d_{34}^2 + (\dd - s_2 d_5)g_{5,s_1, s_2}\right]
 \\
  s_2\frac{i(d_4-id_3)}{d_{34}}d_{12}(\dd - s_2 d_5)
 \\
 d_{34}d_{12}
 },
 \\
 N &=\sqrt{\left(d_{34} g^{(3)}_{s_1, s_2}\right)^2 + \left(d_{34}^2 + (\dd - s_2 d_5)g_{5,s_1, s_2}\right)^2 + \left(d_{12}(\dd - s_2 d_5)\right)^2 + d_{34}^2 d_{12}^2}
 \label{eq:vecs}
\end{aligned}
\end{equation}
 with $s_1, s_2 = \pm 1$, $d_{ij} = \sqrt{d_i^2 + d_j^2}$, $g^{(3)}_{s_1,s_2} =  s_2 B + \dd  + s_1 s_2 \sqrt{B^2 + d^2 + s_2 2 B \dd}$, and
$g_{5, s_1, s_2} = g^{(3)}_{s_1, s_2} - \dd - s_2 d_5$.
The spectrum along the $k_z$ axis is shown near $k_D$ in Fig. \ref{fig:Epmpm}. There are four Weyl nodes at $k_z^{w_1} = k_D$, $k_z^{w_2}=k_D - B/v_+$, $k_z^{w_3} = k_D + B/v_+$, $k_z^{w_4} = k_D$ with energies $E_{w_1} = -B$, $E_{w_2} = -B v_-/v_+$, $E_{w_3} = B v_-/v_+$, $E_{w_4} = B$.

For a Fermi pocket defined by $E_{s_1, s_2} = 0$, the pocket is described by the eigenstate $|\psi_{s_1, s_2}\rangle$ of $\mathcal{H}(\k)$. The Fermi surface projection selects the zero-energy eigenstates of $\mathcal{H}(\k)$ and $\mathcal{H}_h(\k)$, the latter of which has eigenstates 
\begin{equation}
    |\psi^h_{-s_1, s_2}\rangle = -\Pi \mathcal{K}|\psi_{s_1, s_2}\rangle,
\label{eq:psih}
\end{equation}
where $\mathcal{K}$ is the complex conjugation operator, at energy $E^h_{-s_1, s_2} = -E_{s_1, s_2} = 0$.  The overall phase in Eq. \eqref{eq:psih} is chosen for convenience to keep the final component of the state positive. The projected gap function over the $E_{s_1, s_2} = 0$ Fermi pocket to lowest order is then 
\begin{equation}
    \Delta_{proj}(\k) = \langle \psi_{s_1, s_2}| \hat{\Delta}(\k)|\psi^h_{-s_1, s_2}\rangle = - \langle \psi_{s_1, s_2} | \hat{\Delta}(\k) \Pi \mathcal{K} |\psi_{s_1, s_2}\rangle.
\end{equation}

The phase winding in Eq. \eqref{eq:vecs} requires the eigenvectors to be written with different overall phase winding near different Fermi surface points on the $k_z$ axis to avoid phase winding singularities in nonzero components of the eigenvector. For the Fermi pockets with $k_z > 0$, which are topologically nontrivial, the pockets cross the $k_z$ axis twice and we refer to the point on the $k_z$ axis furthest from the $\Gamma$ point ($\k = 0$) as the north pole and the point on the $k_z$ axis closest to the $\Gamma$ point as the south pole. With two such pockets, we call the pocket with the south pole closest to $\Gamma$ $FS_1$ and the other pocket $FS_2$, and the locations of the north and south poles are listed in Table \ref{tab:poles}. The Fermi pockets at $k_z < 0$ are related by parity. When $\mu > \mu_{U,1} \equiv k_D v_p -B$ or $\mu < \mu_{L,1} \equiv -(k_D v_s-B)$, $FS_1$ has merged with the parity-related pocket and is trivial. For $\mu > \mu_{U,2} \equiv k_D v_p +B$ or $\mu < \mu_{L,2} \equiv -(k_D v_s+B)$, there are only two Fermi pockets in the entire Brillouin zone, and both are trivial. The eigenvectors in Eq. \eqref{eq:vecs} can only have phase winding singularities along the $k_z$ axis and can be smoothly defined on any Fermi pocket using at most two patches. The eigenvectors at each Fermi pocket are shown in Table \ref{tab:gauges} in the appropriate smooth gauge at each pole. As in Sec. \ref{app:Chern}, the transition function, and hence the Chern number, can be read from the gauge at each pole together with knowledge of whether the Fermi pocket is electron- or hole-type. For example, at $\mu_{U,1} >\mu > B$, the smaller Fermi pocket, $FS_2$, is electron-type and has $|\psi^N_{++}\rangle = e^{-i\phi_\k}|\psi^S_{++}\rangle$, indicating a transition function with $q=1/2$ in Eq. \eqref{eq:transition} and hence the Chern number $C=+1$. For the hole pocket $FS_2$ at $B\frac{v_-}{v_+} > \mu > -B\frac{v_-}{v_+}$, the appropriate gauges satisfy $|\psi_{--}^N\rangle = e^{2i\phi_\k}|\psi_{--}^S\rangle$, which, due to the hole pocket reversing the normal direction, corresponds to $C=+2$.

\begin{table}[h]
\begin{tabular}{c | c |c | c | c}
$\mu$ & $k_{z,1}^S$ & $k_{z,1}^N$ & $k_{z,2}^S$ & $k_{z,2}^N$ \\
\hline
\hline
$-B > \mu  > \mu_{L,1}$ & $k_D - \frac{B - \mu}{v_s}$ & $k_D + \frac{B - \mu}{v_p}$ & $k_D + \frac{B + \mu}{v_s}$ & $k_D - \frac{B + \mu}{v_p}$ \\
\hline
$-B\frac{v_-}{v_+}> \mu > -B$ & $k_D - \frac{B - \mu}{v_s}$ & $k_D + \frac{B - \mu}{v_p}$ & $k_D - \frac{B + \mu}{v_p}$ & $k_D + \frac{B + \mu}{v_s}$ \\
\hline
$B\frac{v_-}{v_+} > \mu > -B\frac{v_-}{v_+}$ 
& $k_D - \frac{B + \mu}{v_p}$ & $k_D - \frac{B - \mu}{v_s}$  & $k_D + \frac{B+\mu}{v_s}$ & $k_D + \frac{B-\mu}{v_p}$ \\
\hline
$B>\mu>B \frac{v_-}{v_+}$ & $k_D - \frac{B + \mu}{v_p}$ & $k_D + \frac{B + \mu}{v_s}$ & $k_D - \frac{B - \mu}{v_s}$ & $k_D + \frac{B - \mu}{v_p}$ \\
\hline
$\mu_{U,1} > \mu > B$ & $k_D - \frac{\mu+B}{v_p}$ & $k_D + \frac{\mu + B}{v_s}$ & $k_D - \frac{\mu - B}{v_p}$ & $k_D + \frac{\mu - B}{v_s}$
\end{tabular}
\caption{
\label{tab:poles} $k_z$ coordinates for the nodes of the two $k_z>0$ Fermi pockets for $\mu_{U,1}>\mu > \mu_{L,1}$. For larger or smaller $\mu$, $FS_1$ has the north pole $k_{z,1}^N$ given respectively in the first or last row with the remaining pole at $-k_{z,1}^N$. For $\mu > \mu_{U,2}$ or $\mu < \mu_{L,2}$, the poles of $FS_2$ are similarly at $\pm k_{z,2}^N$.}
\end{table}

\begin{table}[h]
\begin{tabular}{c | c |c | c | c}
$\mu$ & $FS_1$ $S$ pole state & $FS_1$ $N$ pole state & $FS_2$ $S$ pole state & $FS_2$ $N$ pole state\\
\hline
\hline
$-B > \mu > \mu_{L,1}$ & $e^{2 i \phi_\k} |\psi_{--}\rangle $ & $e^{3 i \phi_\k} | \psi_{--}\rangle$ &  $e^{i \phi_\k} |\psi_{-+}\rangle $ & $|\psi_{-+}\rangle$\\
\hline
$-B\frac{v_-}{v_+}> \mu > -B$ & $e^{2 i \phi_\k} |\psi_{--}\rangle $ & $e^{3 i \phi_\k}|\psi_{--}\rangle$ & $|\psi_{--}\rangle$ & $e^{i\phi_\k}|\psi_{--}\rangle$ \\
\hline
$B\frac{v_-}{v_+} > \mu > -B\frac{v_-}{v_+}$ &
$|\psi_{+-}\rangle$ & $e^{2i \phi_\k}|\psi_{+-}\rangle$ & $e^{i\phi_\k} |\psi_{--}\rangle$ & $e^{3i \phi_\k}|\psi_{--}\rangle$\\
\hline
$B>\mu>B \frac{v_-}{v_+}$ & $|\psi_{+-}\rangle$ & $e^{i\phi_\k}|\psi_{+-}\rangle$ & $e^{2i\phi_\k} |\psi_{+-}\rangle$ & $e^{3i\phi_\k}|\psi_{+-}\rangle$ \\
\hline
$\mu_{U,1} > \mu > B$ & $|\psi_{+-}\rangle$ & $e^{i\phi_\k}|\psi_{+-}\rangle$ & $e^{3i\phi_\k}|\psi_{++}\rangle$ & $e^{2i\phi_\k}|\psi_{++}\rangle$
\end{tabular}
\caption{
\label{tab:gauges} States describing the two $k_z>0$ Fermi pockets for $\mu_{U,1}>\mu > \mu_{L,1}$ in the appropriate smooth gauge at each pole. }
\end{table}

\newpage 
\section{Monopole superconducting order for different irreps in the model of magnetically doped C\lowercase{d}$_3$A\lowercase{s}$_2$ with $\beta_s = \beta_p$}

The possible pairings in irreducible representations of the $C_{4h}$ point group symmetry of the magnetically doped model in Eq. \eqref{eq:H} can be constructed from products of momentum and matrix irreps in the spin-orbital basis. The representation of the symmetry group acting on the $4\times 4$ spin-orbital basis is generated by $D(i) = I_{2\times 2}\otimes\tau_z$ and $D(C_4) = e^{-i\frac{\pi}{4}[\sigma_z \otimes(2 I_{2\times 2} - \tau_z)]} = \text{diag}[e^{\frac{-i\pi}{4}},e^{\frac{-i3\pi}{4}},e^{\frac{i\pi}{4}},e^{\frac{i3\pi}{4}}]$. Since $C_{4h}$ is abelian, the irreps are one-dimensional, and each matrix $M$ that is in an irrep satisfies $D(g) M D(g)^T = \chi(g) M$ for each group element $g$. Note that the transpose $D(g)^T$ is used rather than the conjugate transpose since the gap function mixes particle and hole blocks of the BdG matrix. In terms of $M_{ij} \equiv \sigma_i \otimes \tau_j$, with $0, 1, 2, 3$ labeling the identity and $x, y,$ and $z$ Pauli matrices, the sixteen matrices in Table \ref{tab:matirreps} span the space of $4\times 4$ matrices and are irreps of $C_{4h}$, 
with the character table reproduced in Table \ref{tab:c4hirreps} for convenience \cite{Bilbao2006}.

\begin{table}[h]
\subfloat[\label{tab:matirreps} Gap function matrix irreps.]{

 \begin{tabular}{c | c}
 Irrep & Matrices \\
 \hline \hline
 $A_g $ & $M_{10}$, $M_{13}$, $M_{20}$, $M_{23}$ \\
 \hline
 $\-^1E_g $ & $M_{00}+M_{33}$, $M_{03}+M_{30}$ \\
 \hline
 $\-^2E_g $ & $M_{00}-M_{33}$, $M_{03}-M_{30}$ \\
 \hline
 $B_u $ & $M_{01}$, $M_{02}$, $M_{31}$, $M_{32}$ \\
 \hline
 $\-^1E_u$ & $M_{11}+M_{22}$, $M_{12}-M_{21}$ \\
 \hline
 $\-^2E_u$ & $M_{11}-M_{22}$, $M_{12}+M_{21}$ \\
 \end{tabular}
}
\hspace{2cm}
\subfloat[\label{tab:c4hirreps} $C_{4h}$ character table.]{
 \begin{tabular}{c|c c c | c}
     Irreps & $E$ & $\Pi$ & $C_4$ & Basis Functions\\
     \hline\hline
     $A_g $ & $1$ & $1$ & $1$ & $k_z^2$, $k_x^2 + k_y^2$ \\
     \hline
     $B_g $ & $1$ & $1$ & $-1$ & $k_x k_y$, $k_x^2 - k_y^2$\\
     \hline
     $\-^1E_g $ & $1$ & $1$ & $-i$ & $k_z(k_x - i k_y)$\\ 
     \hline
     $\-^2E_g $ & $1$ & $1$ & $i$ & $k_z(k_x + i k_y)$\\
     \hline 
     $A_u $ & $1$ & $-1$ & $1$ & $k_z$\\
     \hline 
     $B_u $ & $1$ & $-1$ & $-1$ & $k_z k_x k_y$, $k_z (k_x^2 - k_y^2)$\\
     \hline
     $\-^1E_u $ & $1$ & $-1$ & $-i$ & $k_x - i k_y$\\
     \hline
     $\-^2E_u $ & $1$ & $-1$ & $i$ & $k_x + i k_y$
  \end{tabular}
}
\caption{Irrep tables listing (\textbf{a}) the gap function matrices in each irreducible representation of $C_{4h}$ under the transformation $D(g) M D(g)^T$ and (\textbf{b}) the character table for $C_{4h}$. In (a), The matrices are expressed as tensor products of Pauli matrices, $M_{ij} = \sigma_i \otimes \tau_j$. The character table in (b) includes the characters for the identity $E$ and group generators, parity $\Pi$ and $\pi/2$ clockwise rotation about the $z$ axis $C_4$.}
\end{table}

For any matrix part, the momentum irrep must be chosen so that the condition required by Fermi statistics, $\Delta_{\alpha \beta}(\k) = -\Delta_{\beta \alpha}(-\k)$, is satisfied. Further, the projected gap function vanishes for pairings with matrix part $M_{01}$, $M_{10}$, $M_{11}\pm M_{22}$, $M_{20}$, $M_{23}$, or $M_{31}$.
Note that the projected gap function to lowest order is calculated from $\Delta_{proj}(\k) = \langle \psi_{s_1, s_2}| \hat{\Delta}(\k)|\psi^h_{-s_1, s_2}\rangle = \langle \psi_{s_1, s_2}| \hat{\Delta}(\k) (-\Pi)\mathcal{K}|\psi_{s_1, s_2}\rangle$, with $\Pi = M_{03}$ the parity operator and $\mathcal{K}$ the complex conjugation operator. The projected gap function to lowest order is thus of the form $v^T M v$ and vanishes when the matrix $M = -\hat{\Delta}(\k) \Pi$ is antisymmetric, $M = -M^T$, as is the case for $\hat{\Delta}(\k) \propto M_{01}, M_{11}\pm M_{22}, M_{20}, M_{23}$ and $M_{31}$. These gap functions thus vanish as a result of parity symmetry. The final vanishing gap function, $\hat{\Delta}(\k)\propto M_{10}$, is a special case for the $\beta_s = \beta_p$ model that vanishes since $\psi_{s1, s2}^{(1)}\psi_{s1, s2}^{(3)} - \psi_{s1, s2}^{(2)}\psi_{s1, s2}^{(4)} = 0$, where $\psi_{s1, s2}^{(j)}$ is the $j$th component of $|\psi_{s1, s2}\rangle$ in the spin-orbital basis. In the full model, Eq. \eqref{eq:H} with $\beta_s\neq \beta_p$, we expect a pairing with matrix $M_{10}$ to have qualitatively similar projected gap function structure to one with $M_{13}$, both of which are in the $A_g$ representation.

Accounting for Fermi statistics, the pairings $\hat{\Delta}(\k)$ that lead to nonvanishing projected gap functions can be in one of four irreps, $A_u,$ $B_u,$ $\-^1E_u,$ and $\-^2E_u$.
In the following tables, the phase winding patterns are shown for the nine $M_{ij}$ matrices leading to nonvanishing projected gap functions in the $\beta_s = \beta_p$ model. For simplicity, only momentum-independent pairings and pairings with $k_z/k_D$ momentum dependence are considered. Considering different momentum dependence amounts to simply multiplying by a different function of momentum. The tables include descriptions of the nodes and phase winding from which the vorticity can be read. For rows with multiple pairing matrices listed, the pattern is plotted for the first pairing matrix listed, and the patterns for the rest are qualitatively similar with any differences noted in the description. The projected gap function dispersion near the $k_z$-axis nodes is written in terms of $k_\parallel = \sqrt{k_x^2 + k_y^2}$.

Table \ref{tab:mu0035} lists phase winding patterns for the smaller, topologically trivial electron pocket at $\mu = 35$meV, which is described by the $|\psi_{++}\rangle$ state in Eq. \eqref{eq:vecs}. As this pocket is topologically trivial, $C=0$, the total vorticity vanishes for all pairings. Table \ref{tab:mu002} lists phase winding patterns for the $C = 1$ electron pocket at $k_z > 0$ for $\mu = 20$meV, described by the $|\psi_{++}\rangle$ state. Table \ref{tab:mu0} lists phase winding patterns for the $C = 2$ hole pocket at $k_z > 0$ for $\mu = 0$meV, described by the $|\psi_{--}\rangle$ state. The total vorticity in all cases can be seen to be $2C$. Importantly, the vorticity should be read from the counterclockwise phase winding number with respect to the local normal direction, which points out of the surface shown for electron pockets and into the surface for hole pockets.

\newpage


\begin{longtable}[H] {c c c c  >{\centering\arraybackslash}m{0.25\linewidth} c >{\centering\arraybackslash}m{0.2\linewidth}} 
 \caption{\centering Pairing phase winding and node behavior for the smaller $C=0$ Fermi pocket at $\mu = 35$meV. \label{tab:mu0035}} \\ 
 \toprule 
  Pairing & \ \ & Irrep & \ \ \ & $\Delta_{proj}(\k)$ Phase Winding 
  & \ \ \ \ & $\Delta_{proj}(\k)$ Node Locations \\
  \hline \hline
  \tabc{$\Delta_0 \frac{k_z}{k_D} M_{13}$} & & \tabc{$A_u$} 
  & & \tabc{\vspace{0.2cm}\includegraphics[width=\linewidth]{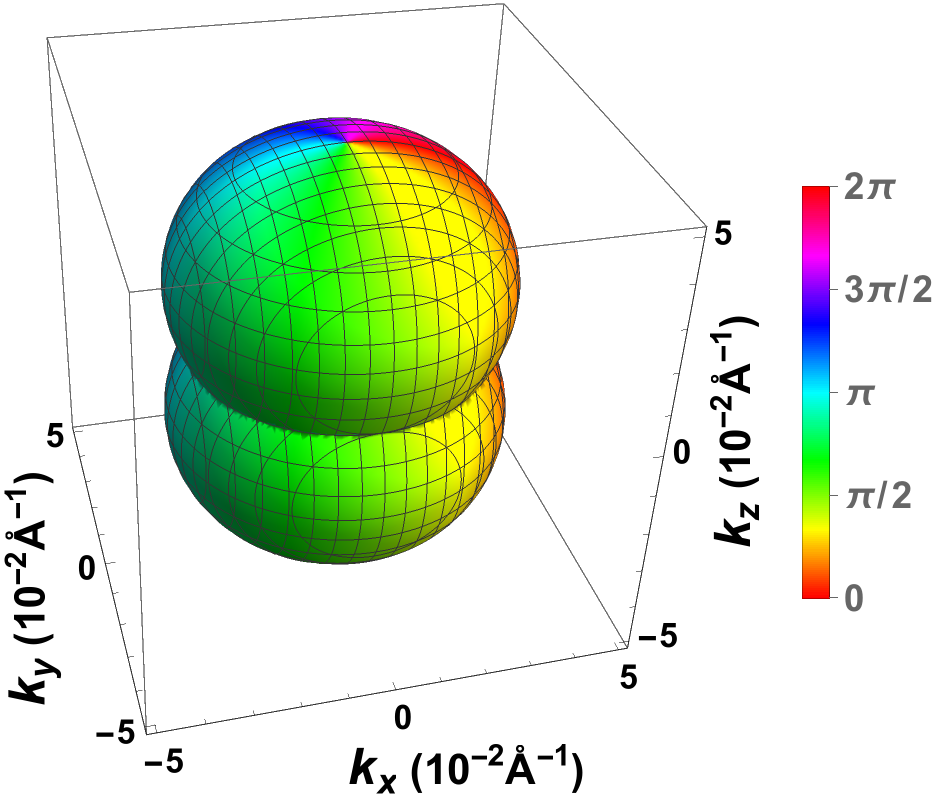}} 
  & &  $k_{\parallel}^3 e^{-i\phi}$ nodes on $k_z$ axis.
  \\
  
  \hline
  \tabc{
      $\Delta_0 M_{02}$ \\
      $\Delta_0 M_{32}$}   
      & & \tabc{$B_u$}  
      & &\tabc{\vspace{0.2cm}\includegraphics[width=\linewidth]{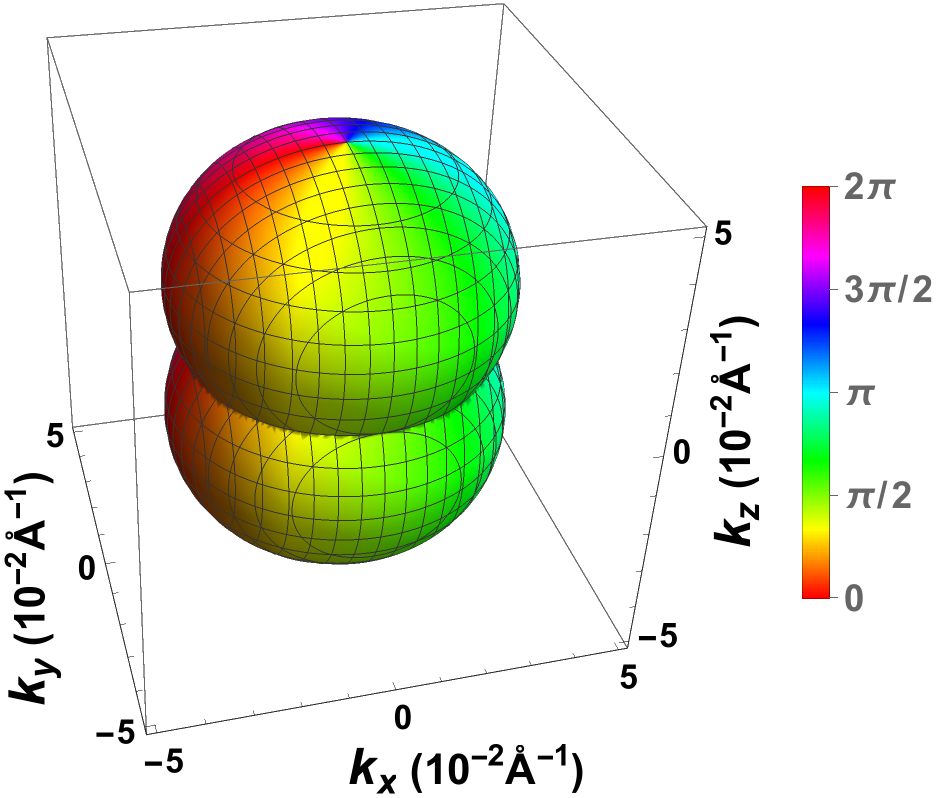}} 
  &\ &  $k_{\parallel} e^{i\phi}$ nodes on $k_z$ axis.
  \\
  
  \hline
  \tabc{
      $\Delta_0 (M_{12} + M_{21})$ \\
      $\Delta_0 \frac{k_z}{k_D} (M_{00} - M_{33})$ \\
      $\Delta_0 \frac{k_z}{k_D} (M_{03} - M_{30})$
      } 
   & &\tabc{$\-^2E_u$} 
   &  & \tabc{\vspace{0.2cm}\includegraphics[width=\linewidth]{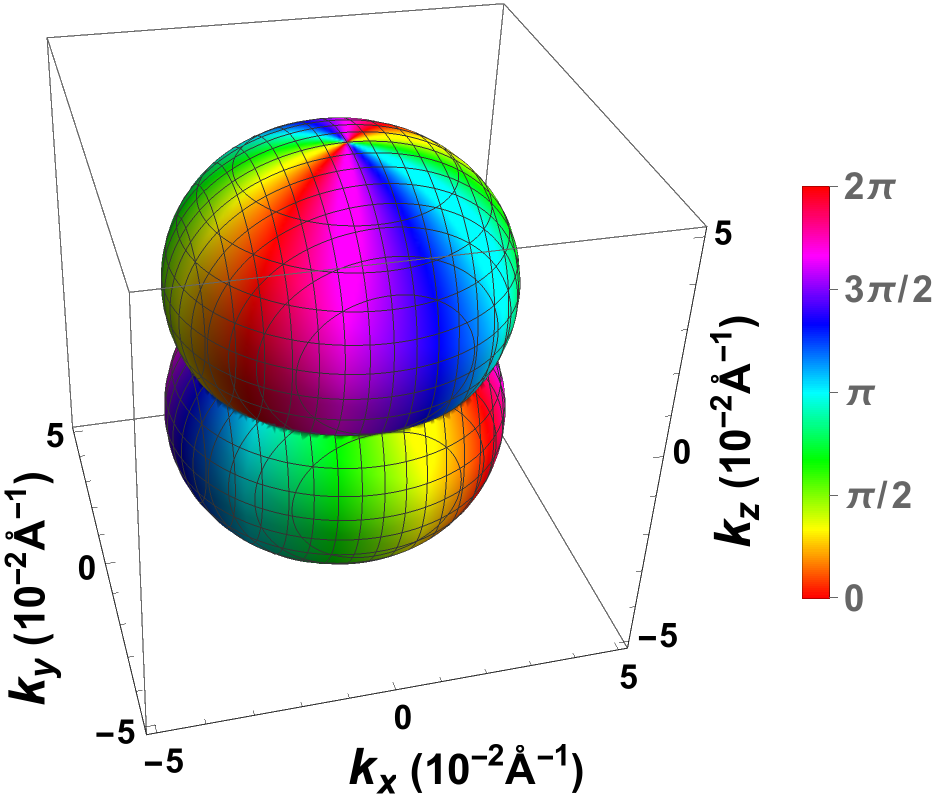}} 
   &\ &  $k_{\parallel}^2 e^{2i\phi}$ nodes on $k_z$ axis.
  \\
  \hline
  \tabc{$\Delta_0 (M_{12} - M_{21})$} 
  & & \tabc{$\-^1E_u$} 
  & &  \tabc{\vspace{0.2cm}\includegraphics[width=\linewidth]{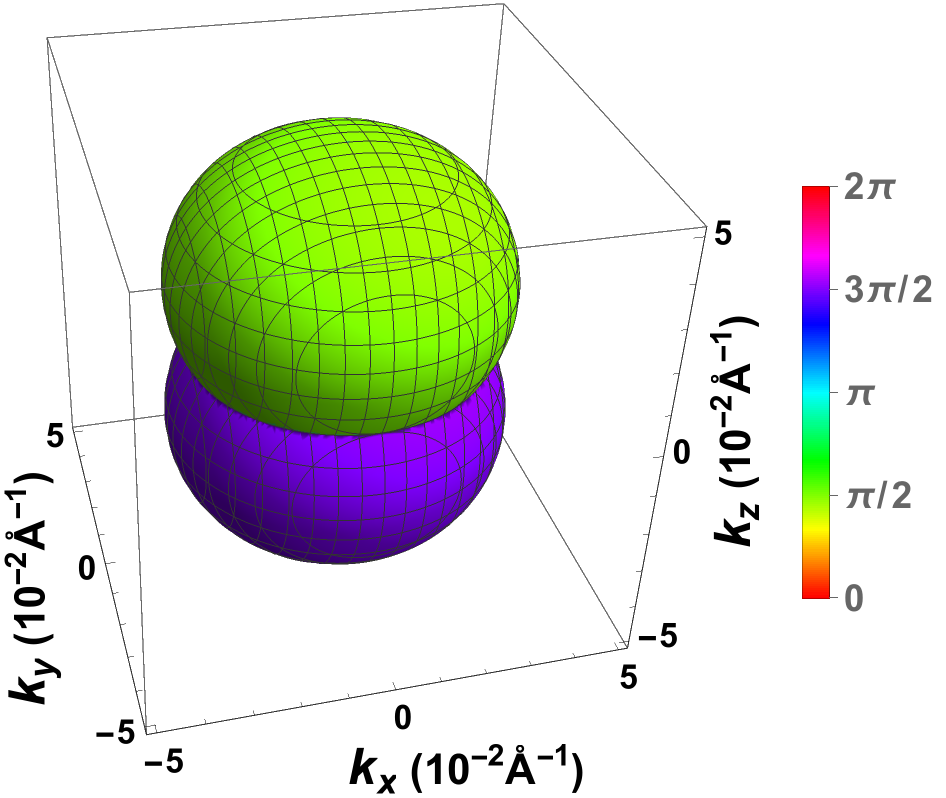}} &  & $k_\parallel^4$ nodes on $k_z$ axis.
  \\
  \hline
  \tabc{$\Delta_0 \frac{k_z}{k_D} (M_{00} + M_{33})$ \\ 
  $\Delta_0 \frac{k_z}{k_D} (M_{03}+M_{30})$} 
  & & \tabc{$\-^1E_u$} 
  & & \tabc{\vspace{0.2cm}\includegraphics[width=\linewidth]{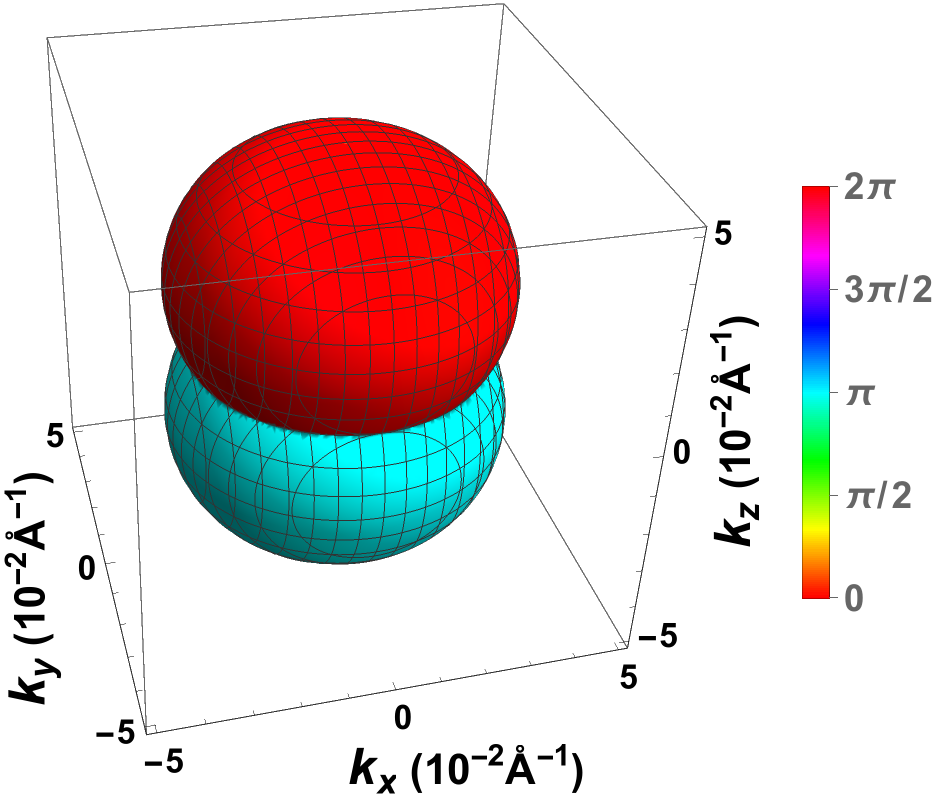}} 
  &  & Gapped on $k_z$ axis. Gap function only vanishes at $k_z = 0$.
  \\ 
  \hline \hline
  
\end{longtable}

\newpage
\begin{longtable}{c c c >{\centering\arraybackslash}m{0.25\linewidth}  >{\centering\arraybackslash}m{0.25\linewidth} c >{\centering\arraybackslash}m{0.15\linewidth}}
\caption{\centering{Pairing phase winding and node behavior for the $C=1$ Fermi pocket at $\mu = 20$meV.} \label{tab:mu002}} \\
    \toprule
    Pairing & Irrep & \ \ \ & $\Delta_{proj}^N(\k)$ Phase Winding (North View) & $\Delta_{proj}^S(\k)$ Phase Winding (South View) & \ \ \ & $\Delta_{proj}(\k)$ Node Locations \\
    \hline \hline
    
    \tabc{$\Delta_0 \frac{k_z}{k_D} M_{13}$} & \tabc{$A_u$} 
    &&  \tabc{\vspace{0.2cm}\includegraphics[width=\linewidth]{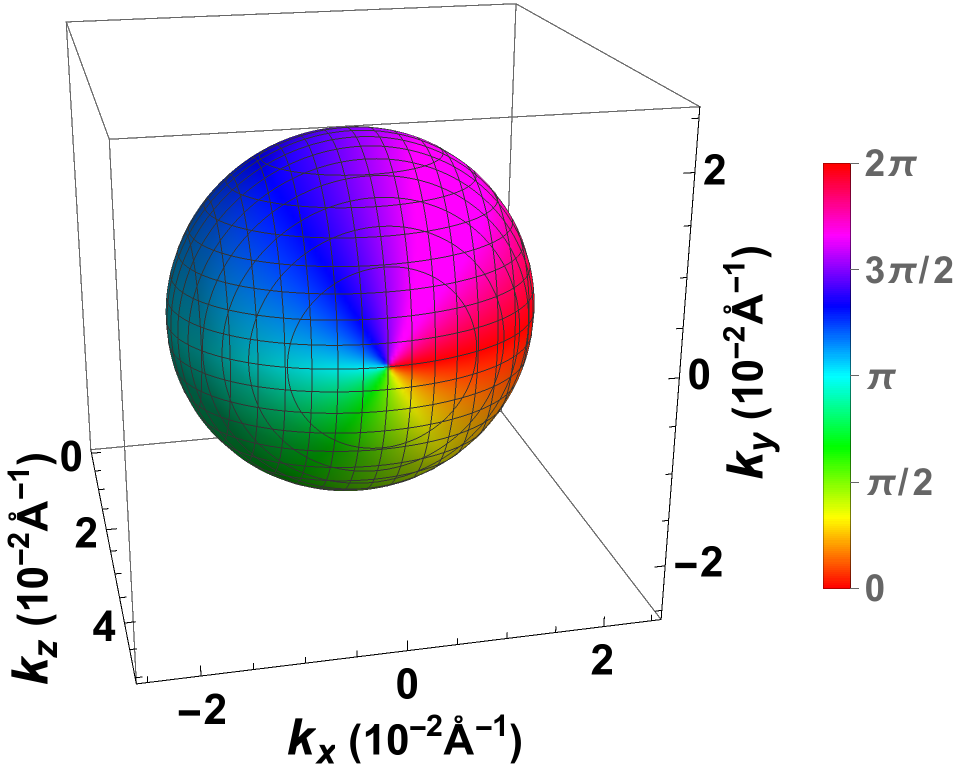}} & \tabc{\vspace{0.2cm}\includegraphics[width=\linewidth]{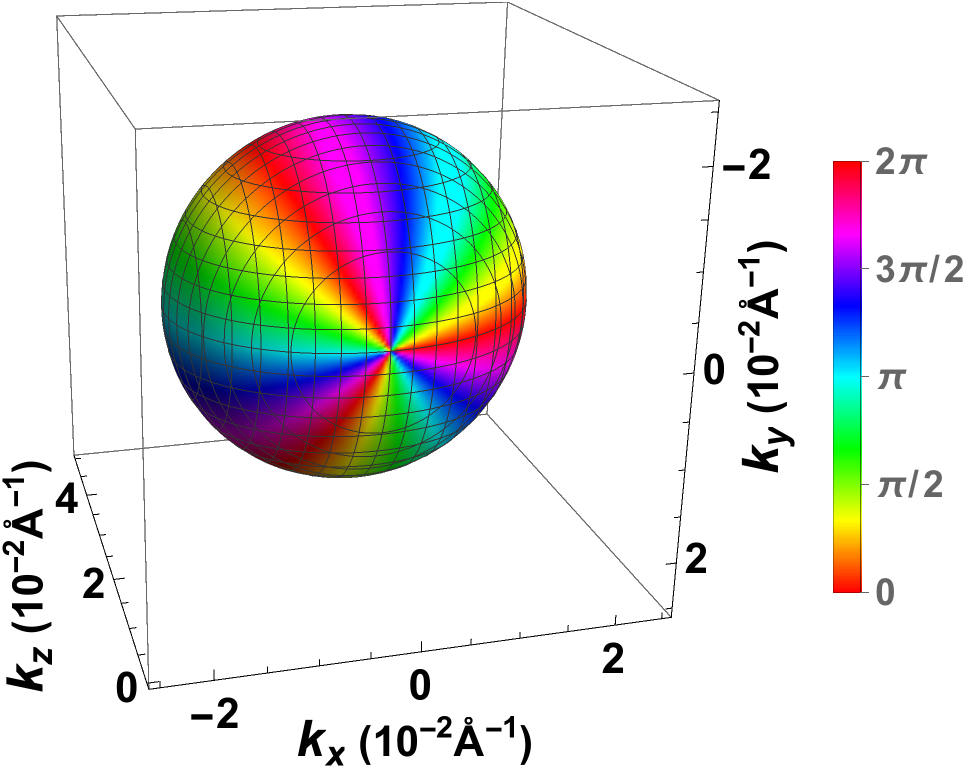}} 
    && $k_{\parallel}^3 e^{-i\phi}$ node at $N$, $k_{\parallel}^3 e^{-3i\phi}$ node at $S$.
    \\
    
    \hline
    \tabc{$\Delta_0 M_{02}$ \\
    $\Delta_0 M_{32}$} & $B_u$ 
    &&  \tabc{\vspace{0.2cm}\includegraphics[width=\linewidth]{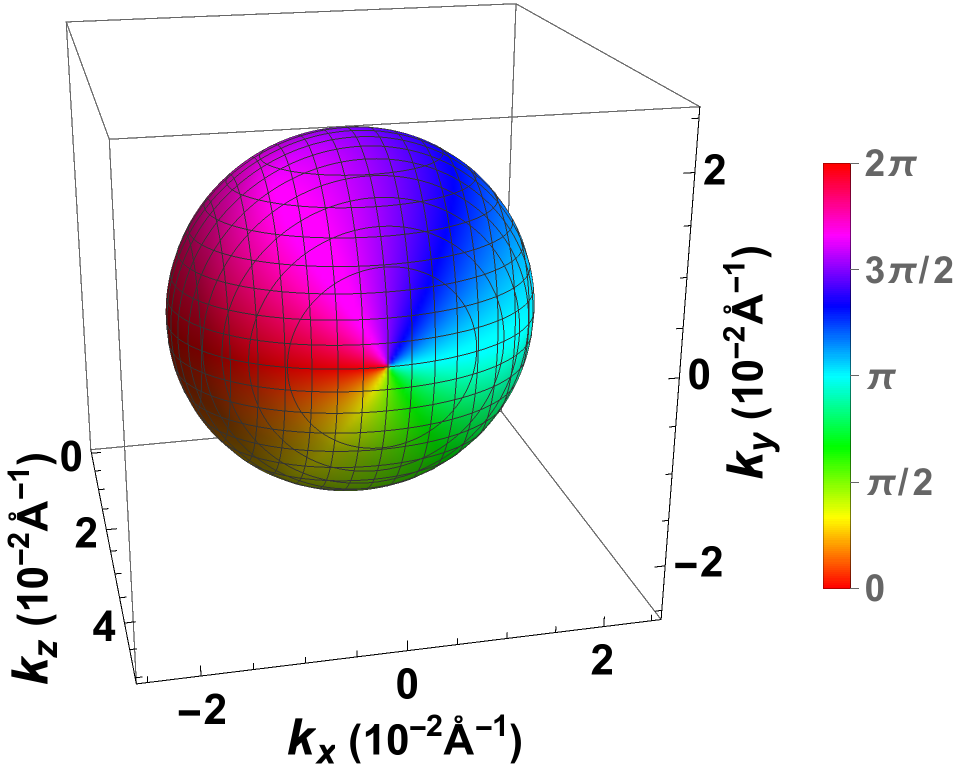}} & \tabc{\vspace{0.2cm}\includegraphics[width=\linewidth]{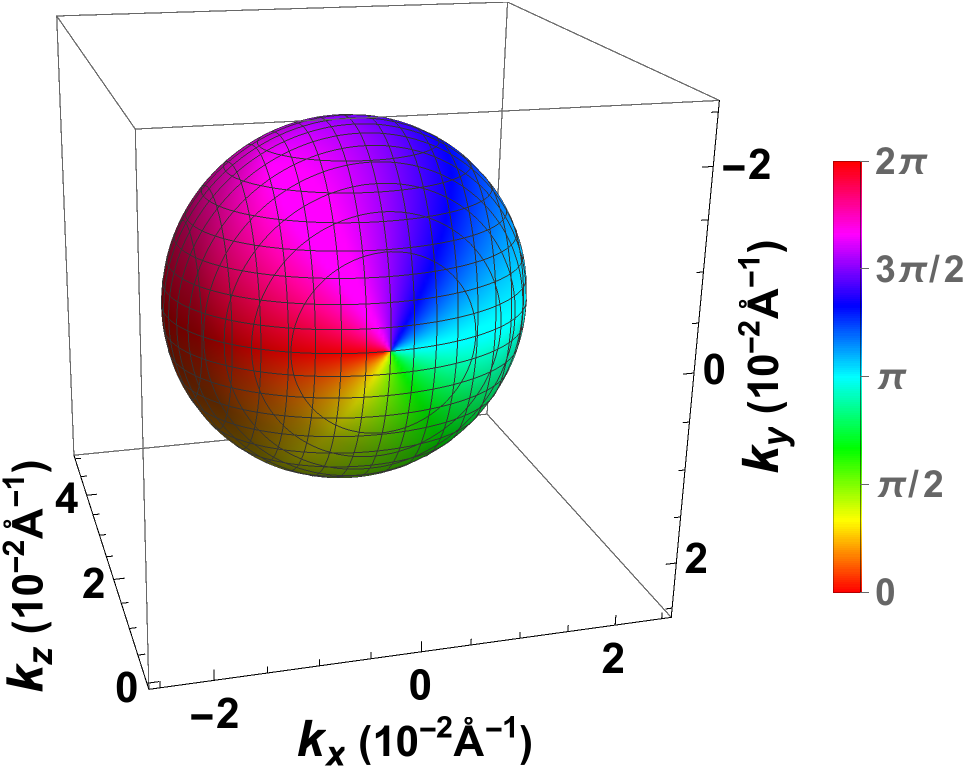}} 
    && $k_{\parallel} e^{i\phi}$ node at $N$, $k_{\parallel} e^{-i\phi}$ node at $S$.
    \\
    
    \hline
    \tabc{$\Delta_0 (M_{12} + M_{21})$} & \tabc{$\-^2E_u$} 
    &&   \tabc{\vspace{0.2cm}\includegraphics[width=\linewidth]{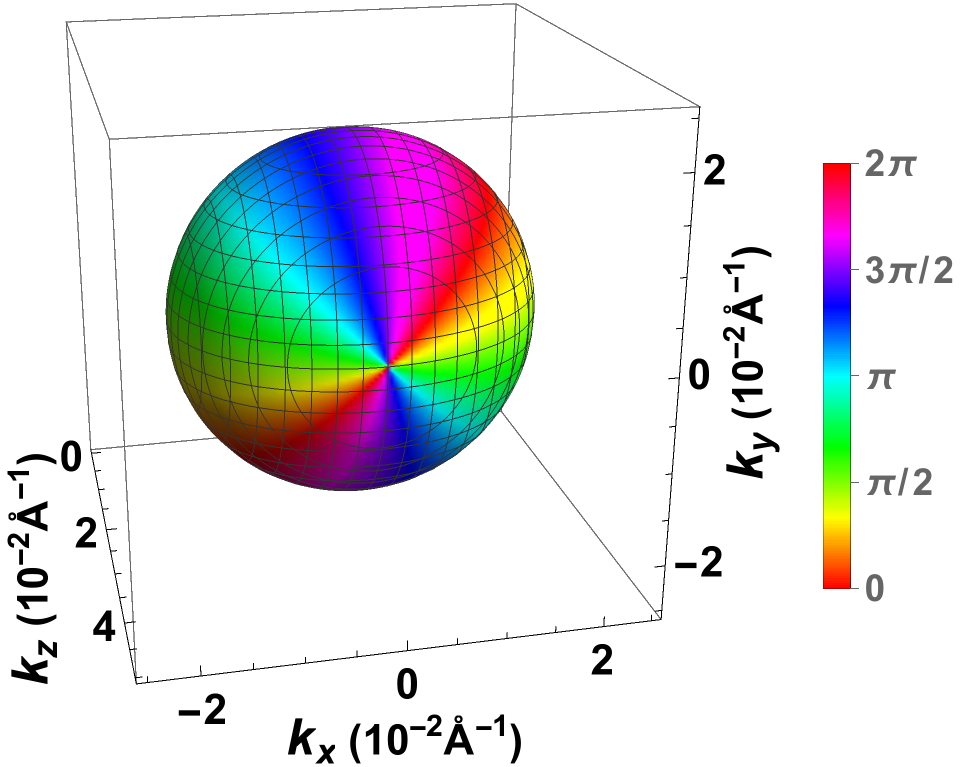}} & \tabc{\vspace{0.2cm}\includegraphics[width=\linewidth]{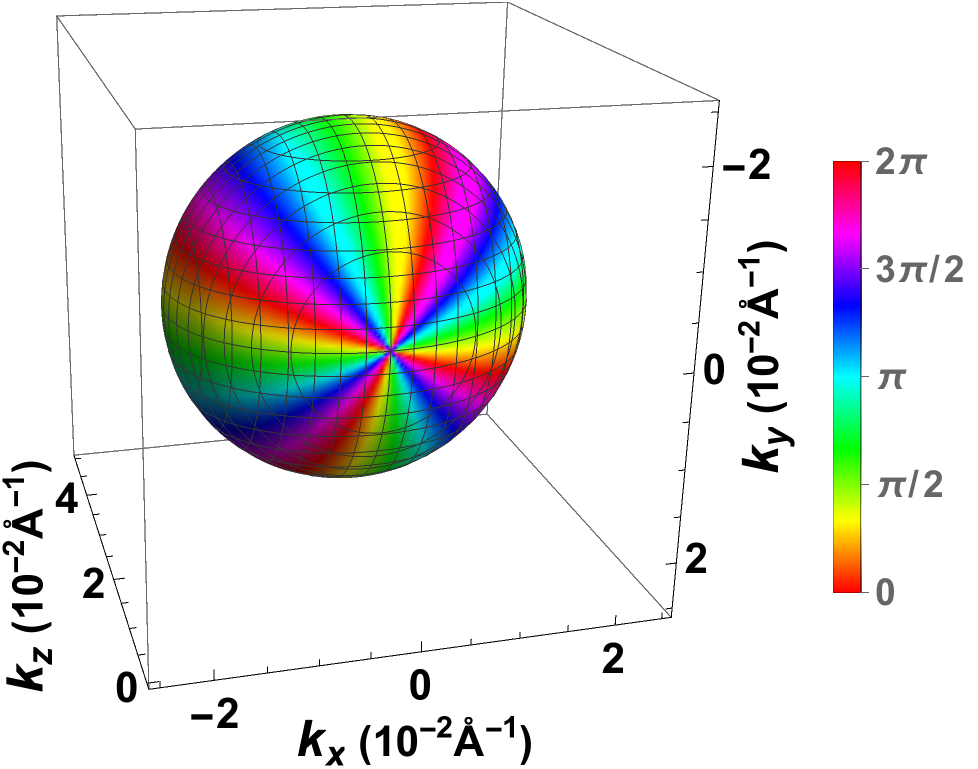}} 
    && $k_\parallel^2 e^{-2i\phi}$ node at $N$, $k_\parallel^4 e^{-4i\phi}$ node at $S$.
    \\
    
    \hline
    \tabc{$\Delta_0 \frac{k_z}{k_D} (M_{00} - M_{33})$ \\
    $\Delta_0 \frac{k_z}{k_D} (M_{03} - M_{30})$} & \tabc{$\-^2E_u$} 
    &&  \tabc{\vspace{0.2cm}\includegraphics[width=\linewidth]{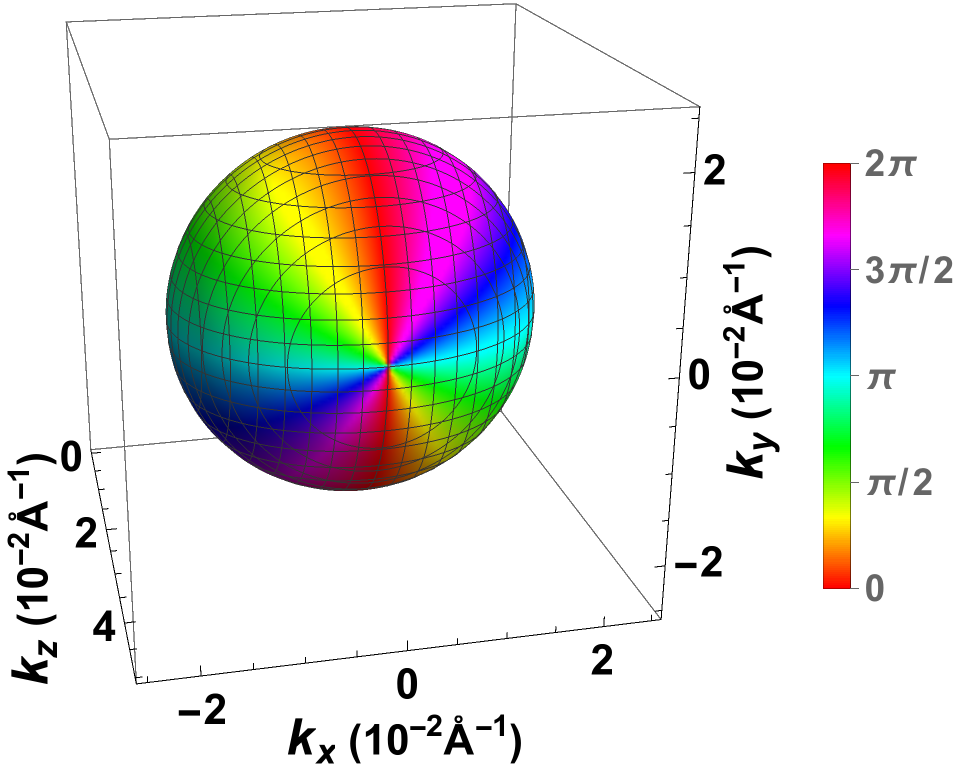}} & \tabc{\vspace{0.2cm}\includegraphics[width=\linewidth]{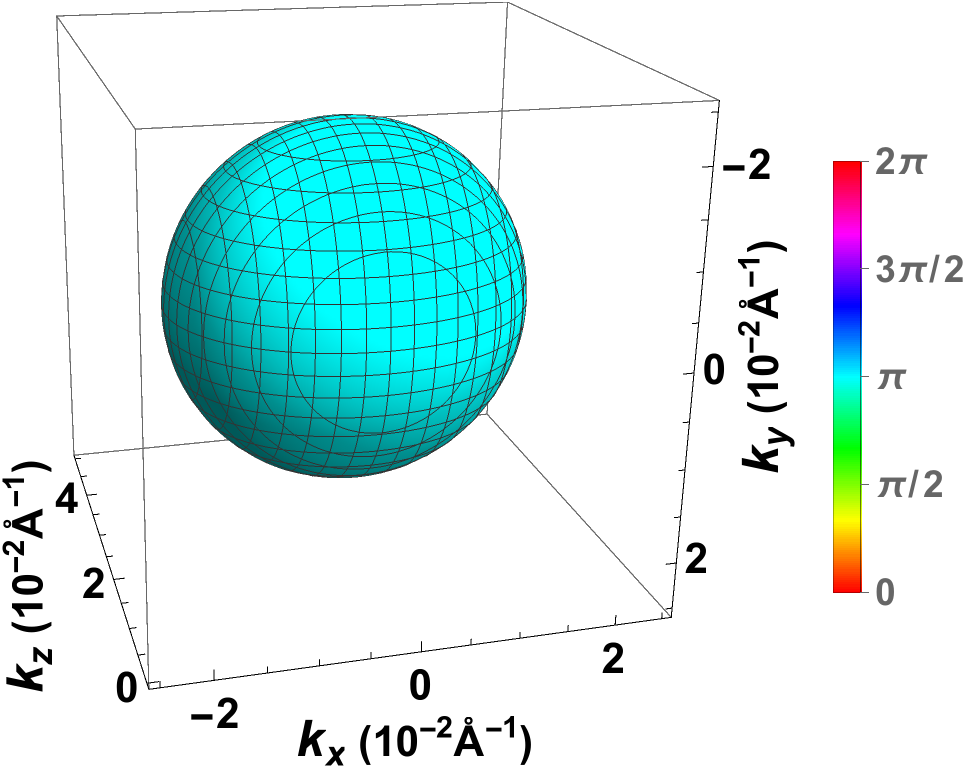}}
    &&  $k_{\parallel}^2 e^{2i\phi}$ node at $N$, gapped at $S$.
    \\

    \hline
    \tabc{$\Delta_0 (M_{12} - M_{21})$} & \tabc{$\-^1E_u$} 
    &&   \tabc{\vspace{0.2cm}\includegraphics[width=\linewidth]{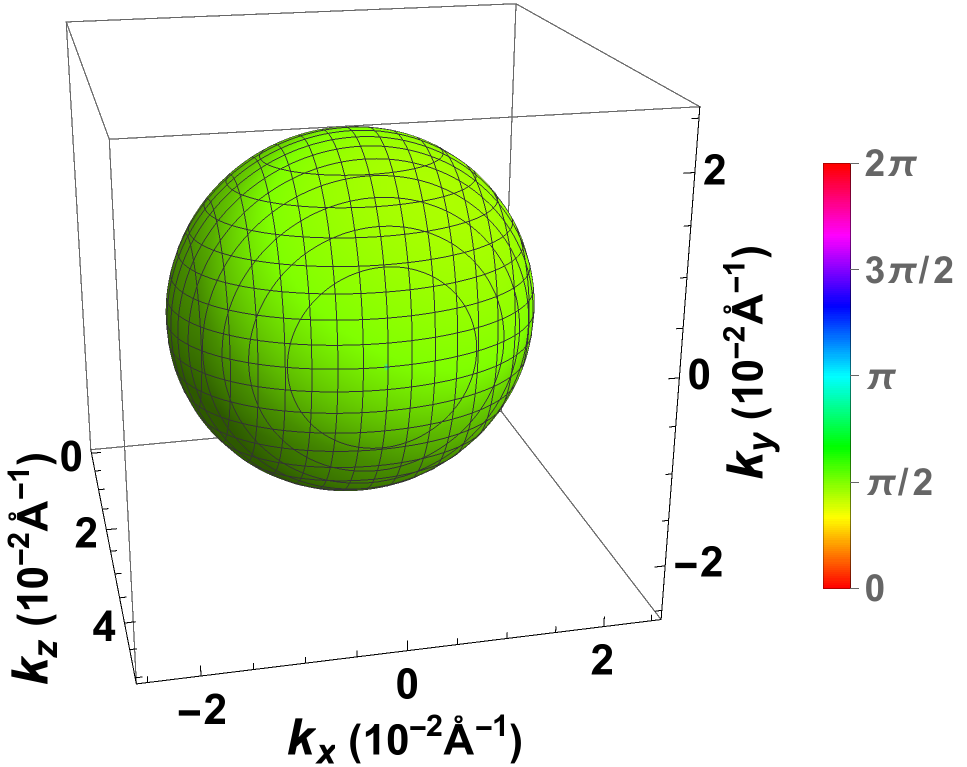}} & \tabc{\vspace{0.2cm}\includegraphics[width=\linewidth]{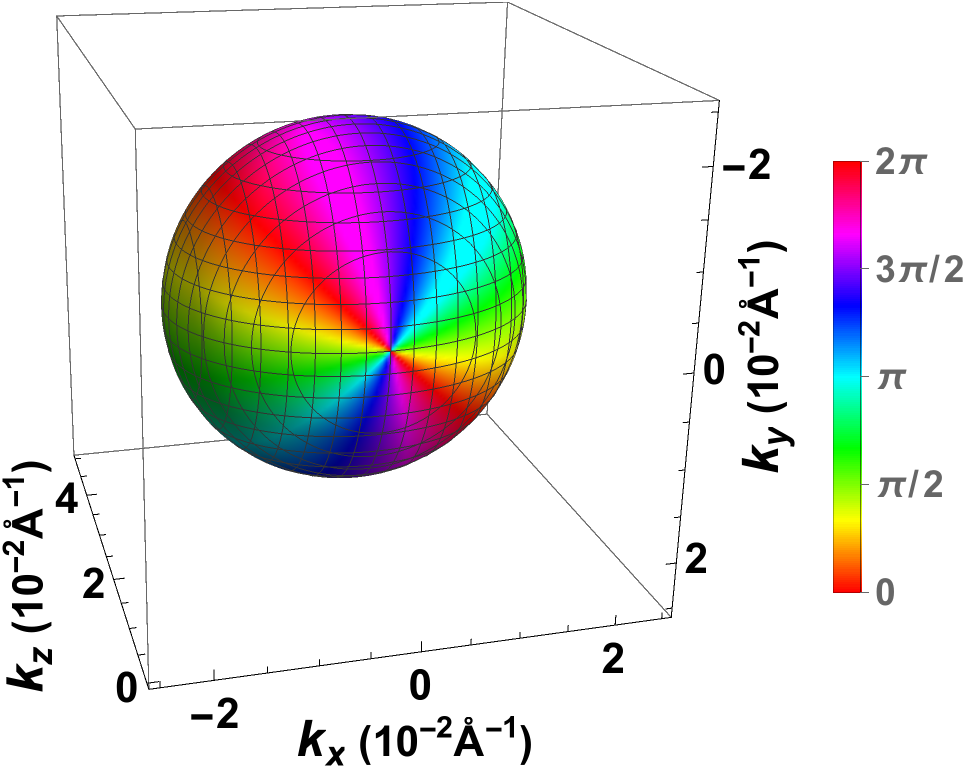}} 
    && $k_\parallel^4$ node at $N$, $k_\parallel^2 e^{-2i\phi}$ node at $S$.
    \\
    
    \hline
    \tabc{$\Delta_0 \frac{k_z}{k_D} (M_{00} + M_{33})$ \\
    $\Delta_0 \frac{k_z}{k_D} (M_{03} + M_{30})$} & \tabc{$\-^1E_u$} 
    &&  \tabc{\vspace{0.2cm}\includegraphics[width=\linewidth]{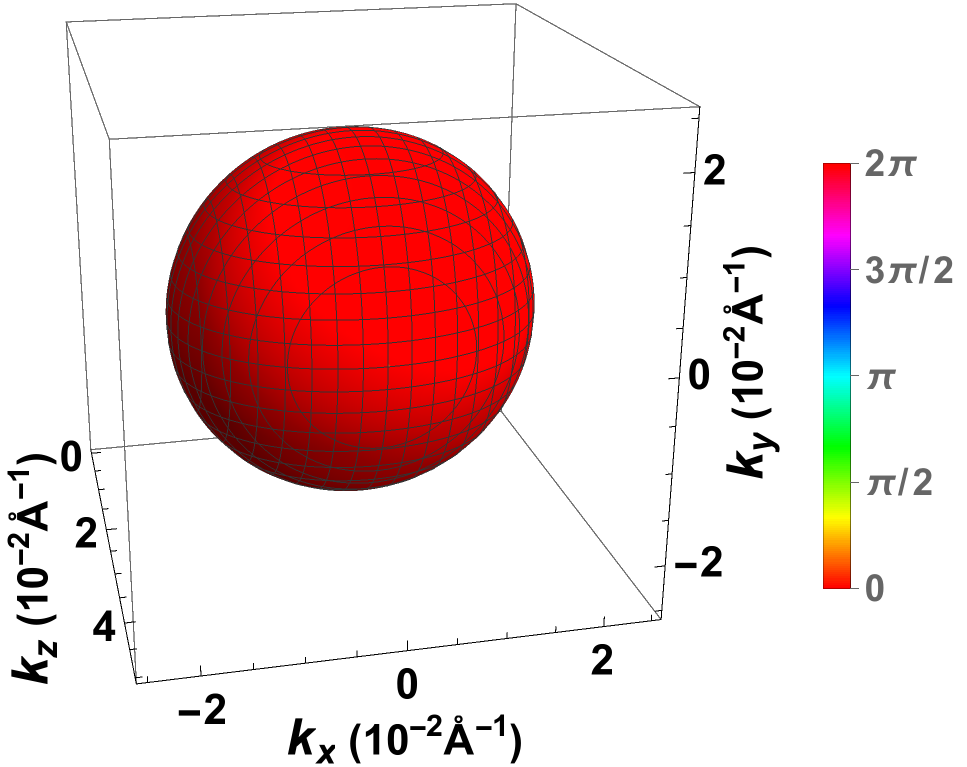}} & \tabc{\vspace{0.2cm}\includegraphics[width=\linewidth]{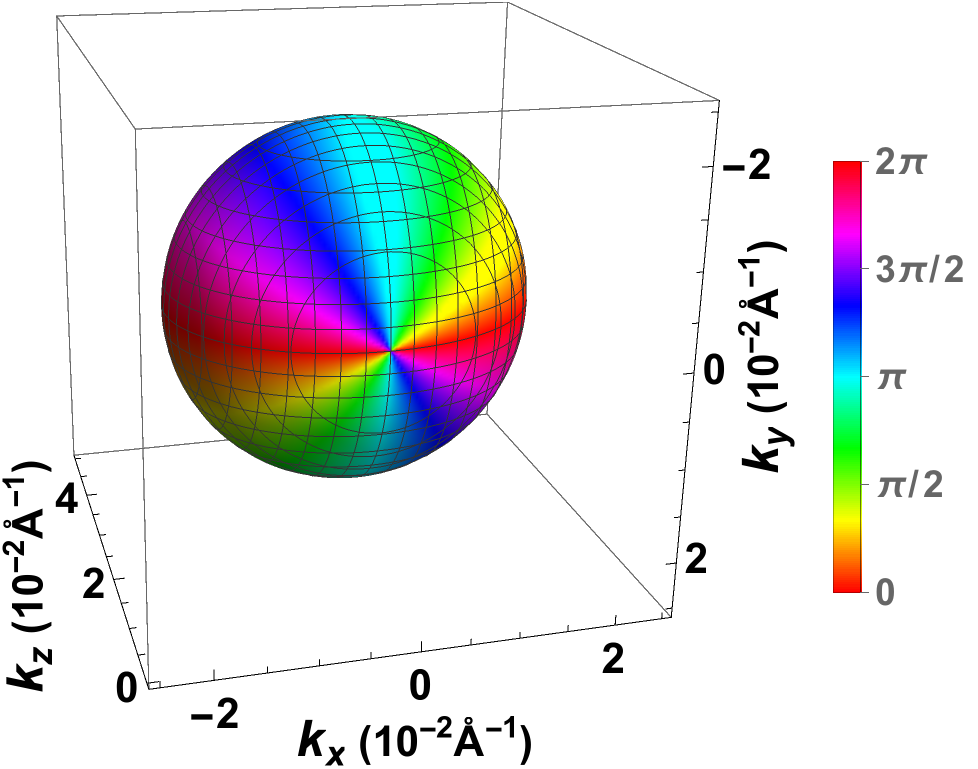}} 
    && Gapped at $N$, $k_{\parallel}^2 e^{-2i\phi}$ node at $S$.
    \\
    \hline \hline
    
\end{longtable}


\begin{longtable}
{>{\centering\arraybackslash}m{0.20\linewidth} c  >{\centering\arraybackslash}m{0.25\linewidth}  >{\centering\arraybackslash}m{0.25\linewidth}  >{\centering\arraybackslash}m{0.25\linewidth}}
\caption{\centering Pairing phase winding and node behavior for the 
$C=2$ hole pocket at $\mu = 0$ meV.\label{tab:mu0}}
\\
\toprule
 Pairing & Irrep & $\Delta_{proj}^N(\k)$ Phase Winding (North View) & $\Delta_{proj}^S(\k)$ Phase Winding (South View) & $\Delta_{proj}(\k)$ Node Locations \\
  \hline \hline
    \tabc{$\Delta_0 \frac{k_z}{k_D} M_{13}$} & \tabc{$A_u$} &  \tabc{\vspace{0.2cm}\includegraphics[width=\linewidth]{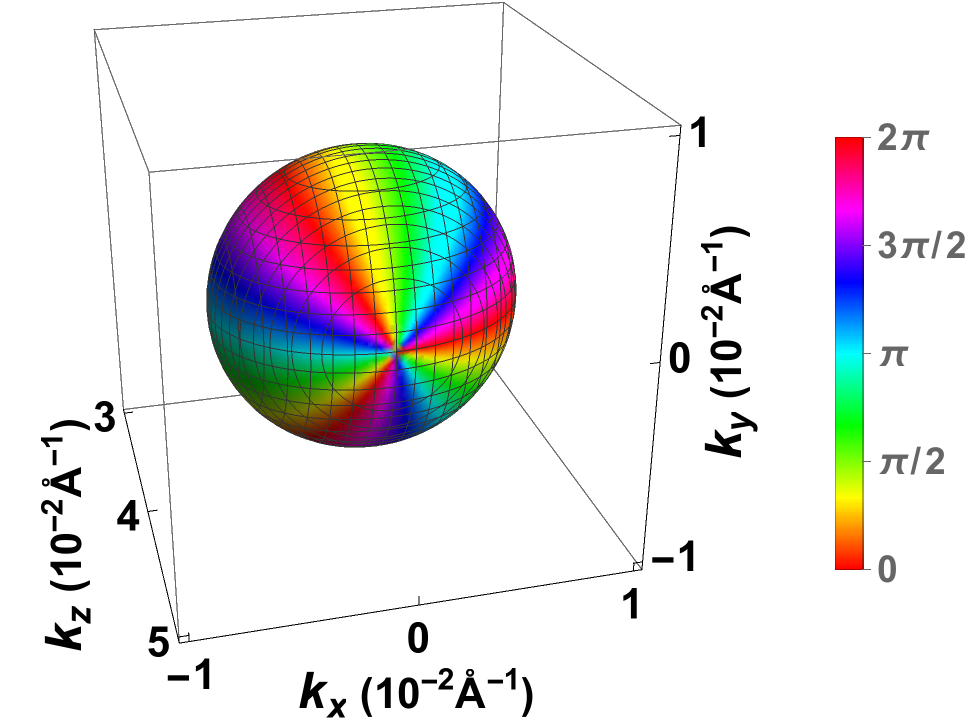}} & \tabc{\vspace{0.2cm}\includegraphics[width=\linewidth]{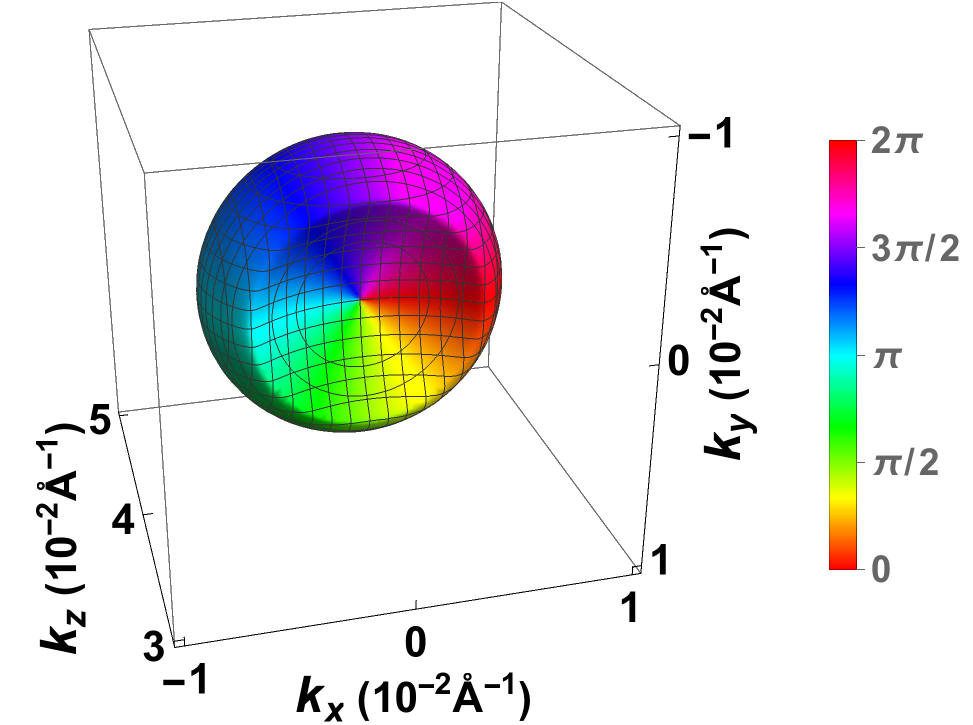}}
    & 
    $k_{\parallel}^3 e^{-3i\phi}$ node at $N$, $k_{\parallel}^3 e^{i\phi}$ node at $S$.
    \\
    
    \hline
    \tabc{$\Delta_0 M_{02}$ \\    $\Delta_0 M_{32}$} 
    & \tabc{$B_u$} 
    & \tabc{\vspace{0.2cm}\includegraphics[width=\linewidth]{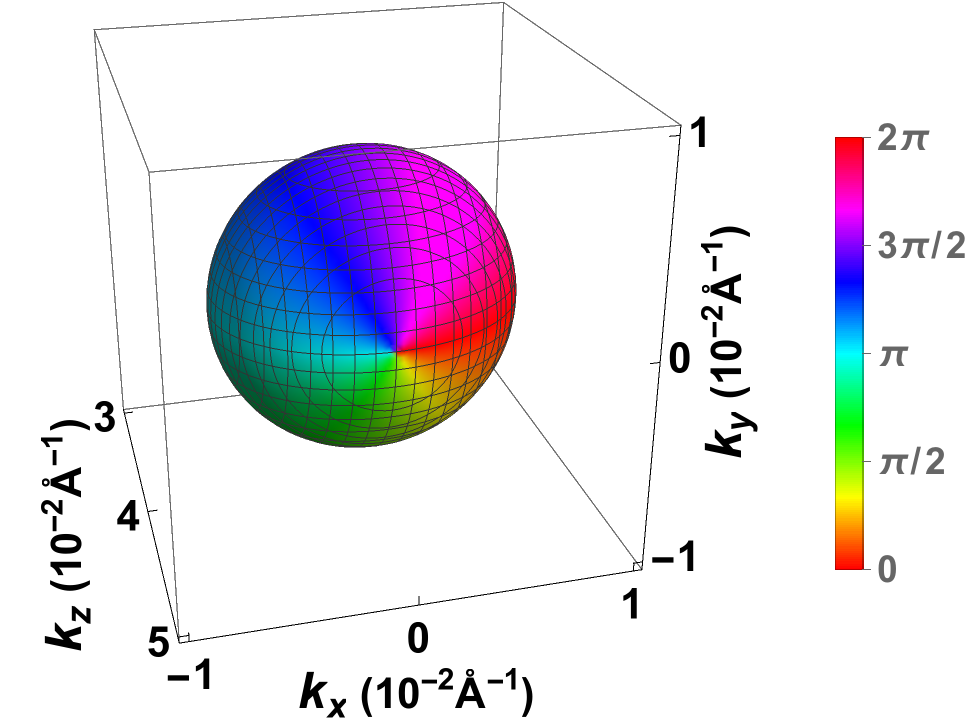}} & \tabc{\vspace{0.2cm}\includegraphics[width=\linewidth]{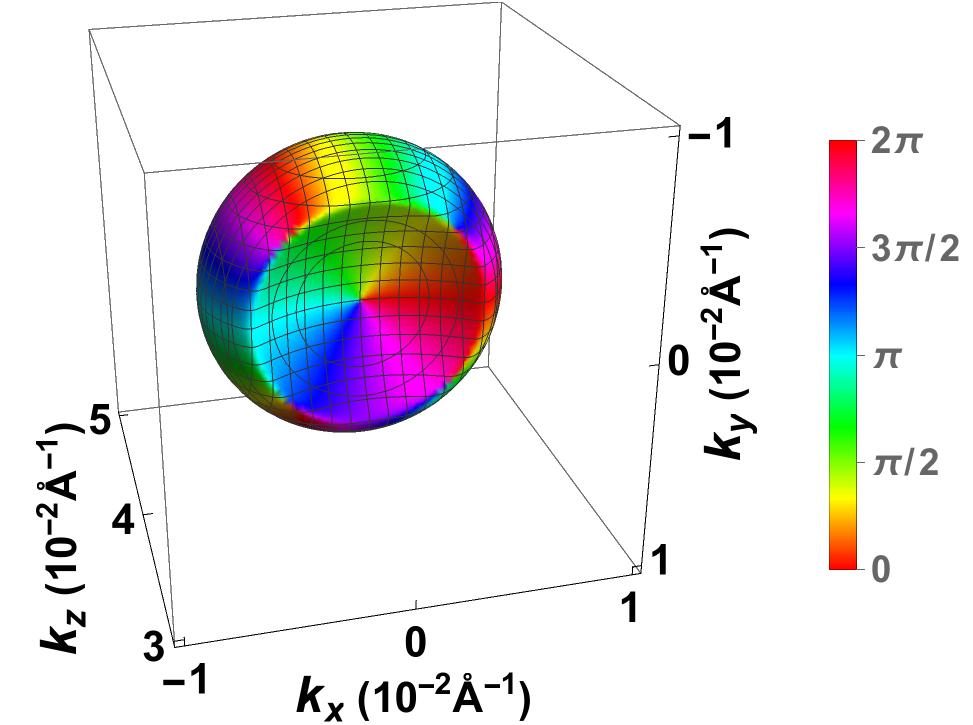}} 
    & {$k_\parallel e^{-i\phi}$ nodes at N and S poles. Four nodes with $e^{i \phi'}$ local winding at $\phi = \pm \pi/4, \pm 3\pi/4$ around concave region near S.
    For $M_{32}$, these nodes are at $\phi = 0, \pm \pi/2, \pi$ instead.}
    \\

    \hline
    \tabc{$\Delta_0 (M_{12} + M_{21})$} & \tabc{$\-^2E_u$} & \tabc{\vspace{0.2cm}\includegraphics[width=\linewidth]{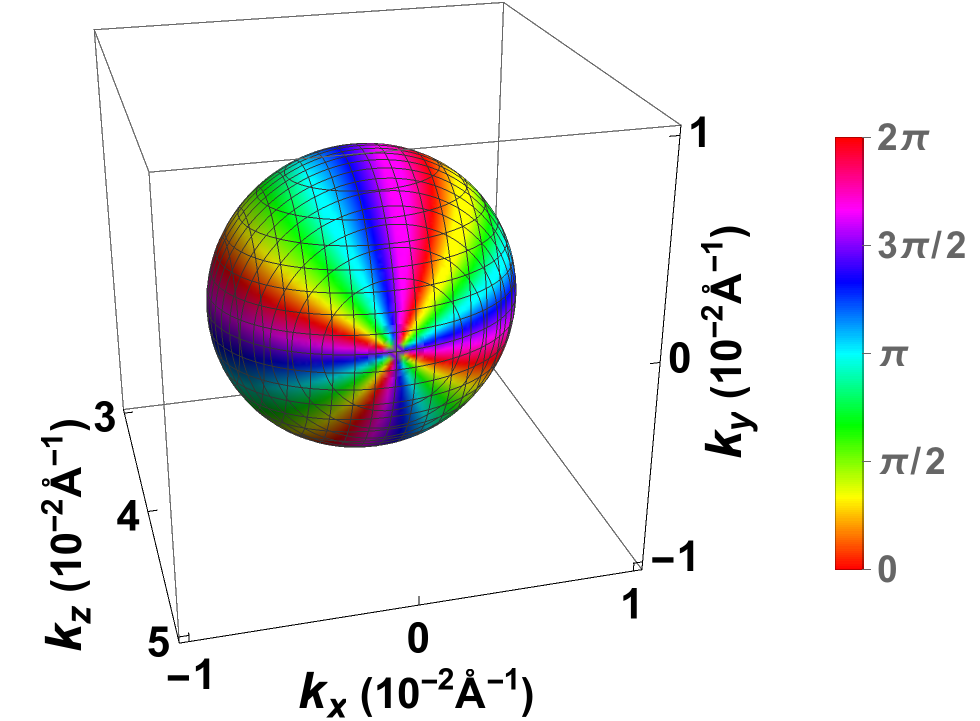}} & \tabc{\vspace{0.2cm}\includegraphics[width=\linewidth]{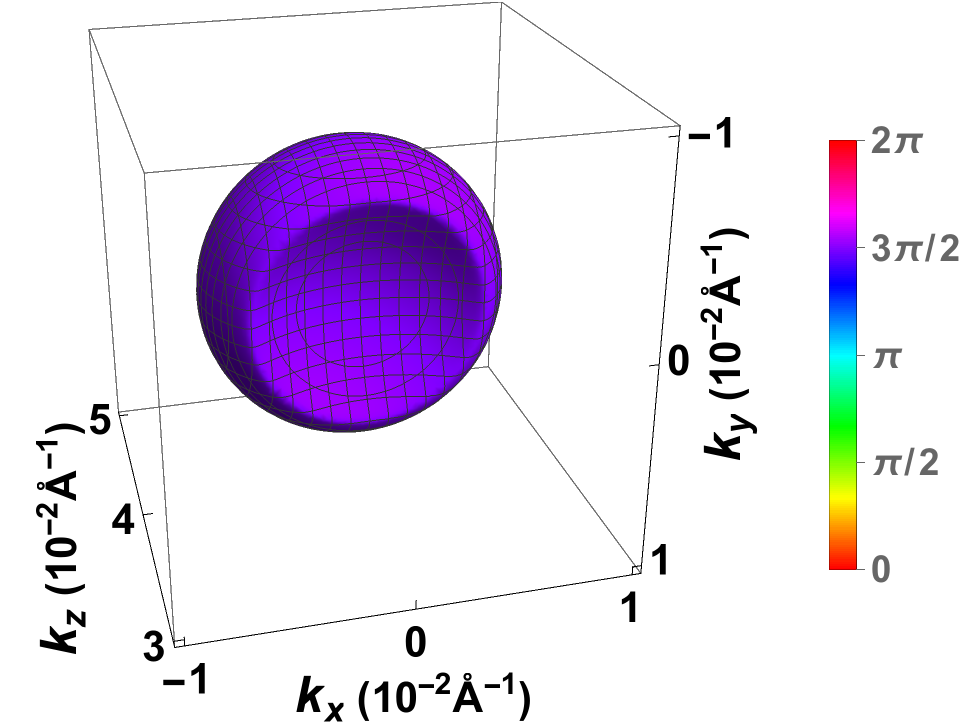}} & $k_\parallel^4 e^{-4i\phi}$ node at $N$, $k_\parallel^4$ node at $S$.
    \\
    
    \hline
    \tabc{$\Delta_0 \frac{k_z}{k_D} (M_{00} - M_{33})$\\
    $\Delta_0 \frac{k_z}{k_D} (M_{03} - M_{30})$} & \tabc{$\-^2E_u$} &  \tabc{\vspace{0.2cm}\includegraphics[width=\linewidth]{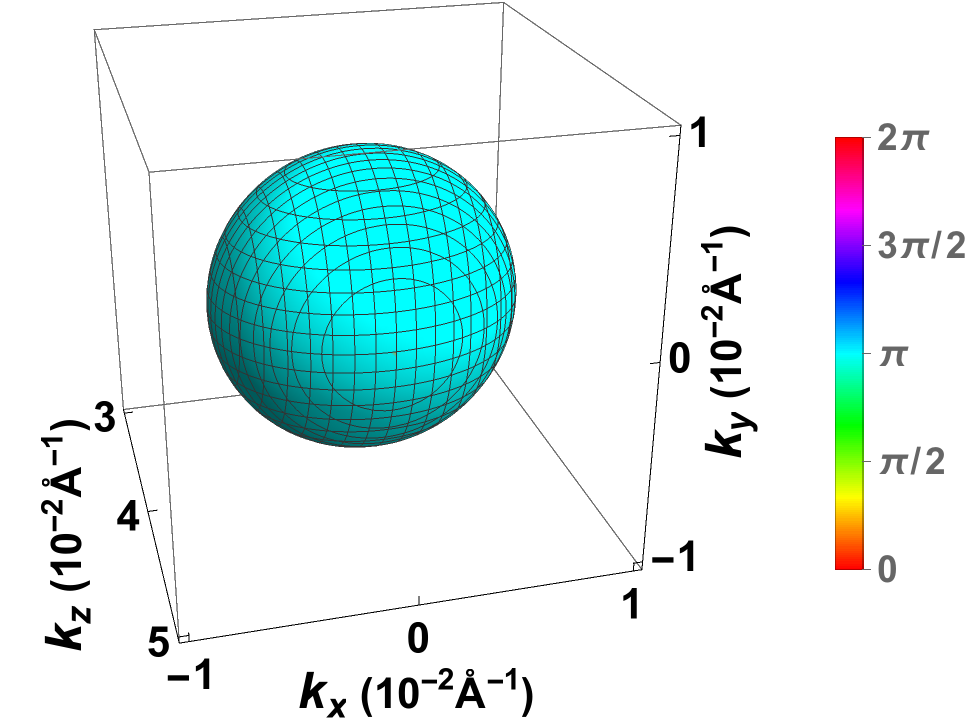}} & \tabc{\vspace{0.2cm}\includegraphics[width=\linewidth]{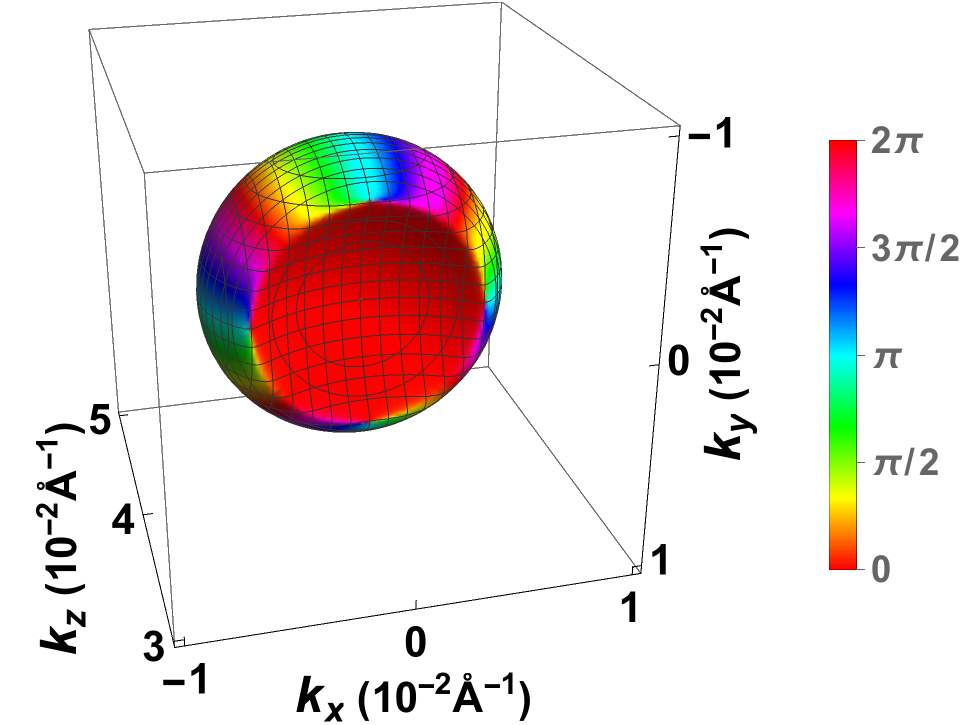}} & 
    Gapped at N and S. Four nodes with $e^{i \phi'}$ local winding at $\phi = 0, \pm \pi/2, \pi$ around the concave region near S.
    For $M_{03}-M_{30}$, these nodes are at $\phi = \pm \pi/4, \pm 3\pi/4$ instead.
    \\

    \hline
    \tabc{$\Delta_0 (M_{12} - M_{21})$} & \tabc{$\-^1E_u$} &  \tabc{\vspace{0.2cm}\includegraphics[width=\linewidth]{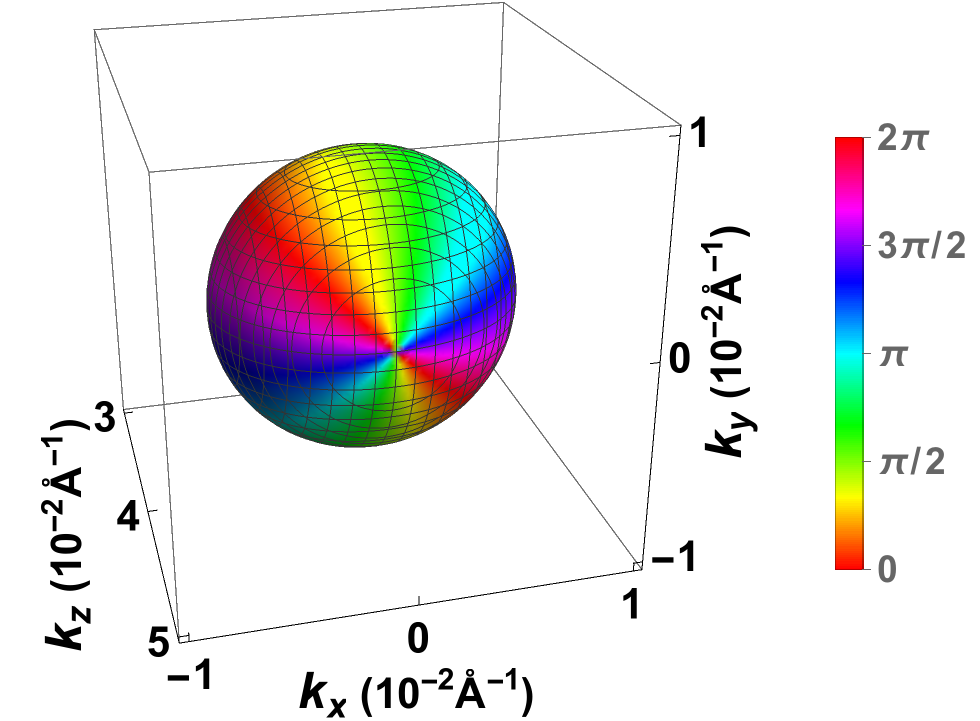}} & \tabc{\vspace{0.2cm}\includegraphics[width=\linewidth]{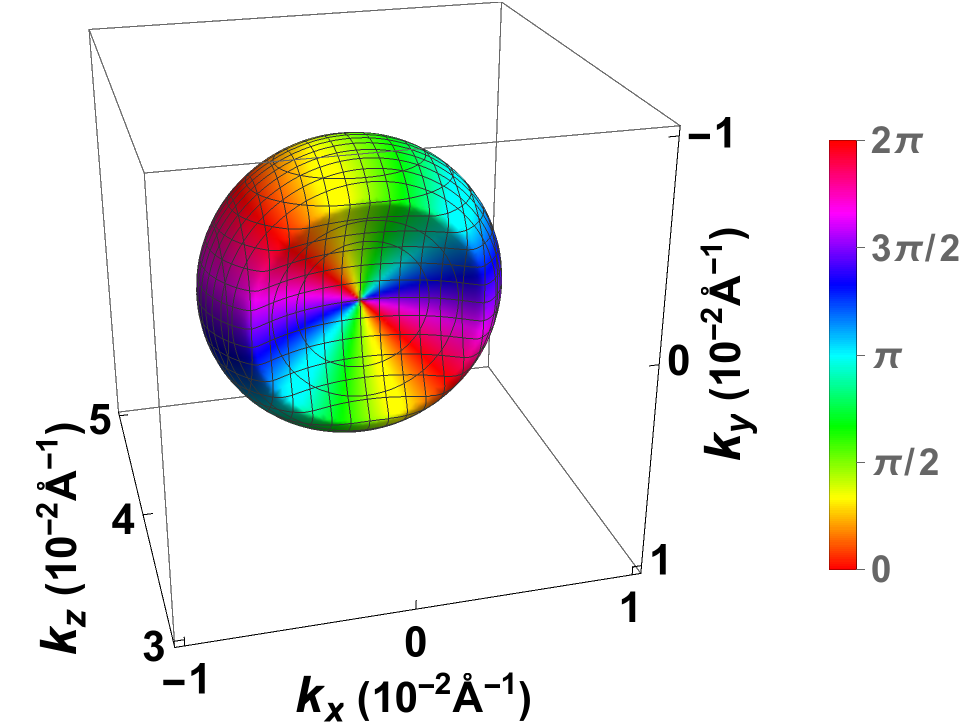}} & 
    $k_\parallel^2 e^{-2i\phi}$ node at $N$, $k_\parallel^2 e^{2i\phi}$ node at $S$.
    \\
    
    \hline
    \tabc{$\Delta_0 \frac{k_z}{k_D} (M_{00} + M_{33})$\\
    $\Delta_0 \frac{k_z}{k_D} (M_{03} + M_{30})$} & \tabc{$\-^1E_u$} &  \tabc{\vspace{0.2cm}\includegraphics[width=\linewidth]{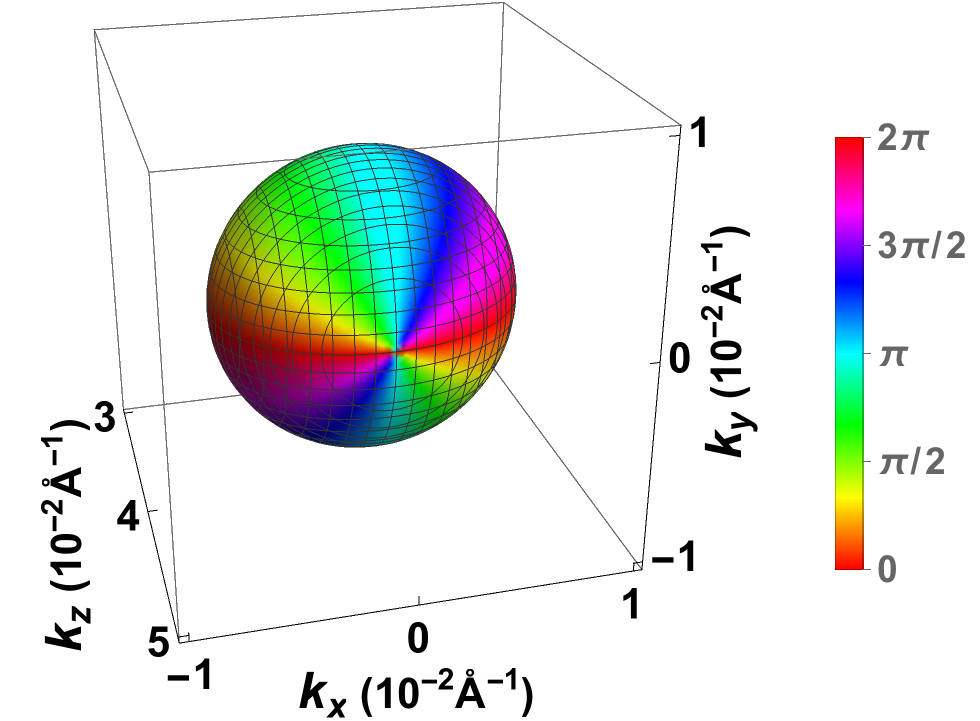}} & \tabc{\vspace{0.2cm}\includegraphics[width=\linewidth]{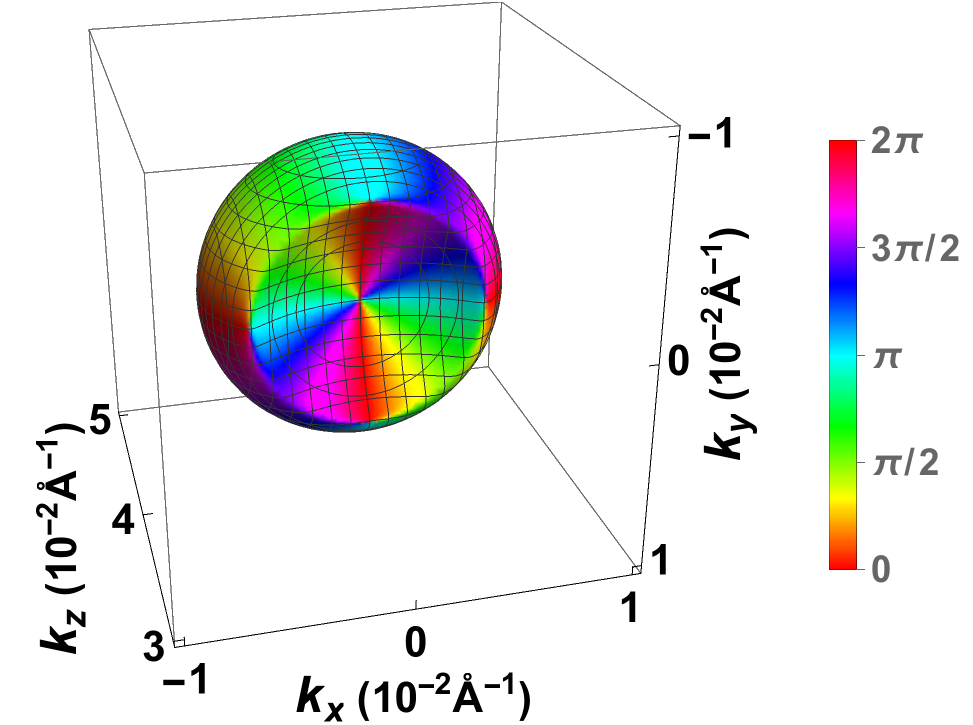}} &
    $k_\parallel^2 e^{-2i\phi}$ nodes at N and S poles. Four nodes with $e^{i \phi'}$ local winding at $\phi = 0, \pm \pi/2, \pi$ around concave region near S. For $M_{03}-M_{30}$, these nodes are at $\phi = \pm \pi/4, \pm 3\pi/4$ instead.
    \\
    \hline
    \hline

\end{longtable}


%